\documentclass{JFM-FLM_Au}

\usepackage{graphicx}
\usepackage{epstopdf, epsfig}
\usepackage{amsmath}
\usepackage{amssymb}
\usepackage{color}
\usepackage{subfigure}
\usepackage{upgreek}
\usepackage{url}

\usepackage[makeroom]{cancel}
\usepackage{qtree}
\usepackage[normalem]{ulem}

\usepackage[resetlabels]{multibib}
\newcites{Supp}{References}

{\vskip 2pt \begin{compactitem}[#1]\setlength{\itemsep}{2pt}}
{\end{compactitem}\vskip 2pt}

\newcommand{\be}{\begin{equation}}
\newcommand{\ee}{\end{equation}}

\newcommand{\bu}{{\bf u}}

\newcommand{\bx}{{\bf x}}

\newcommand{\bomega}{{\mbox{\boldmath $\omega$}}}

\lefttitle{D. Zhao, X. Huang, and G. Li}
\righttitle{Journal of Fluid Mechanics}

\title{Interfacial dynamics and energy cascade in immiscible Rayleigh-Taylor turbulence}

\author{Dongxiao Zhao\aff{1},
 Xiaoxue Huang\aff{2},
 \and Gaojin Li\aff{1}
 }

\affiliation{\aff{1}School of Ocean and Civil Engineering, Shanghai Jiao Tong University, Shanghai 200240, PR China
\aff{2}Department of Physics, National University of Defense Technology, Changsha, Hunan 410000, PR China}

\corresau{Gaojin Li, gaojinli@sjtu.edu.cn}

\begin{document}
\maketitle

\begin{abstract}

We investigate interfacial dynamics and multiscale energy transfer in immiscible Rayleigh–Taylor turbulence using numerical simulations with varying surface tension coefficients $\sigma$. Capillarity is shown to control characteristic length scales, interfacial area, and global energy and enstrophy budgets. The flow exhibits self-similar evolution with respect to surface tension, with the maximum kinetic energy scaling as $\sigma^{1/2}$ and the flow duration as $\sigma^{-1/4}$. A scale-by-scale budget shows that surface tension removes kinetic energy at large scales while injecting it at small scales, with the crossover occurring near the Hinze scale. We further recast and verify a local kinematic relationship between surface-tension power and interface stretching, up to conservative transport, $ \boldsymbol{f}^\sigma\cdot\bu = - \sigma\mathcal{S}~|\nabla c| + \mathrm{Transport}$, where $\boldsymbol{f}^{\sigma}$ is the surface-tension force, $\boldsymbol{u}$ the velocity, $c$ the heavy-fluid volume fraction, and $\mathcal{S}$ the interface stretch rate. This relation links kinetic-energy transfer to the scalar-variance cascade and shows that energy transfer to the interface is governed by local strain. Statistics of individual bubbles and droplets reveal vertically elongated filaments with diameters of about three capillary scales, yielding a linear volume-area relation. Their vertical velocities scale with the square root of equivalent diameter, consistent with drag-buoyancy balance. These findings, particularly the direct link between surface tension power and resolved interface stretching, provide a rigorous physical framework for developing subgrid-scale closures for large eddy simulation of immiscible turbulent flows.
\end{abstract}


\section{Introduction}
When a heavy fluid overlies a lighter fluid in a gravitational field, the interface becomes unstable to the Rayleigh–Taylor (RT) instability \citep{Rayleigh83,Taylor50}. In this process, small interfacial perturbations grow due to baroclinic effects arising from the misalignment of the density and pressure gradients. The RT instability evolves from an initial linear stage, where perturbation amplitudes are small relative to their wavelengths and individual modes grow independently, to a nonlinear saturation stage in which mode coupling leads to the formation of coherent bubbles and spikes \citep{Zhou17-1,Zhou17-2}. Eventually, the system enters self-similar turbulent stages, characterized by a mixing width that scales quadratically with time and the development of a fully established energy cascade \citep{Zhouetal2021Review,Zhou_2024book}.  
RT instability is of broad relevance in both natural and engineering contexts. Examples include astrophysical flows such as supernova explosions and remnants \citep{arnett1989supernova,blondin2001rayleigh}, accretion disk dynamics \citep{krumholz2009formation}, degradation of ignition performance in inertial confinement fusion \citep{ZhouARFM2025}, mixing processes in combustion chambers \citep{Sykes2021ProCI}, and the spread of underwater oil spills \citep{Brizzolara24PNAS}.  

In practical scenarios, RT instability is frequently complicated by additional physical mechanisms, including compressibility, mass ablation, rotation, and magnetic fields \citep{Xu_Zhao_2025JFM,Briard2024JFM,BianZhao26PoP}. More recently, much attention has shifted towards immiscible RT flows where interfacial dynamics and surface tension play a dominant role \citep{Chertkov05PRE,Young06JoT,Zanella20POF,Tavares21PTRS,ZhaoLi26IJMF}. These multiphase systems are ubiquitous in nature and industry, ranging from upper-ocean mixing to the processing of food emulsions, pharmaceuticals, and cosmetics \citep{Mcclements10AnnRevFood}.
In such regimes, particularly when the fluids possess comparable densities, gravitational acceleration drives the interface into a turbulent emulsion characterized by a broad spectrum of spatial and temporal scales. It has been well-established that, within the context of homogeneous isotropic turbulence, the size distribution of polydisperse droplets in the breakup-dominated regime follows a $-10/3$ power-law scaling \citep{DeaneStokes02Nature}. Beyond isotropic turbulence, this scaling universality has been corroborated across various flow configurations, including channel flows \citep{Lu25JFM}, Taylor–Couette turbulence \citep{WangSun22JFM}, and even yield stress fluids \citep{Girotto22JoT}. However, the extent to which this classical scaling applies to the non-homogeneous, anisotropic, gravity-driven RT turbulence remains an open question. 


Regarding energy cascade in immiscible RT turbulence, \cite{Chertkov05PRE} developed a phenomenological model extending prior miscible RT theories \citep{Zhou17-1,Zhou17-2}. Their analysis confirmed quadratic growth of the turbulent mixing zone ($L \propto t^2$) and a Kolmogorov cascade between $L$ and the viscous scale $\eta \propto t^{-1/4}$. Surface tension produces an emulsion-like state with characteristic droplet size $l_d \propto t^{-2/5}$, determined by the balance between kinetic and interfacial energy densities. For $l_d \geq \eta$, scales above $l_d$ follow an Obukhov–Corrsin cascade, while scales below $l_d$ develop a capillary-wave energy cascade along droplet surfaces in parallel with bulk Kolmogorov turbulence. At late times, droplet size shrinks below viscous scale and the capillary cascade collapses, yielding a fine emulsion characterized by Batchelor-type velocity fields and logarithmic density structure functions.  
The above 3-D phenomenological theory has been extended to 2-D immiscible RT flows using lattice Boltzmann simulations \citep{Tavares21PTRS}. Within the adapted Bolgiano–Obukhov framework, large-scale mixing in 2-D flows retains the 3-D scalings ($L \propto t^2$, $U \propto t$). However, unlike the 3-D droplet-size scaling ($l_d \propto t^{-2/5}$), 2-D droplets scale as $l_d \propto t^{2/11}$ , with total interface length $L_{\text{tot}} \propto t^{20/11}$ and enstrophy growth $\Omega \propto t^{3/2}$. These findings indicate that dimensionality modifies scaling exponents while preserving the core physics of immsicible RT turbulence. 

Recently, \cite{Brizzolara24PNAS} validated this phenomenological theory in immiscible Boussinesq RT turbulence via experiments and direct numerical simulations of immiscible fluids, such as oil–water systems. They confirmed the predicted $L \propto t^2$ and $l_d \propto t^{-2/5}$ scalings in \cite{Chertkov05PRE}, along with the coexistence of bulk gravity-driven and sub-droplet-scale capillary-driven turbulence. Their key advancement was identifying a single, time-independent control parameter, $Re_0$, that quantifies the scale separation between the mean droplet size and the Kolmogorov scale and thus governs the existence of capillary-driven turbulence.
This capillary-driven turbulence regime emerges only when $Re_0 > 1$, i.e., when the Kolmogorov scale is smaller than the mean droplet size. The existence of this regime enables prediction of the temporal evolution of interfacial area-to-volume ratios, a key quantity for assessing oil-spill biodegradation efficiency.

Phenomenological models for immiscible RT turbulence predict the temporal scalings of key length scales and global energies, but the underlying inter-scale transfer mechanisms are more directly revealed by scale-by-scale budgets. In multiphase turbulence, such analyses have been developed through several complementary approaches, including spectral, coarse-grained, wavelet-based, and point-splitting formulations. Spectral methods have been used to quantify kinetic-energy spectra and fluxes in binary-fluid turbulence and emulsions \citep{Perlekar19JFM,Crialesi22JFM,Crialesi23CP}. Meanwhile, coarse-grained formulations provide a physical-space description of scale-to-scale transfer in flows with spatially varying density, viscosity, or phase distribution \citep{Aluie13}. Related wavelet-based methods similarly combine scale and spatial localization \citep{FreundFerrante2019}. From a complementary point-splitting perspective, \citet{Thiesset20JFM} introduced a scale-space description of liquid transport, separating the transport of liquid volume fraction in physical space from transport across scales. More recently, \citet{ThiessetVahe25JFM} derived a Kármán-Howarth-Monin equation for multiphase turbulence, accounting for density and viscosity variations across phases as well as surface-tension effects, thereby providing a direct scale-by-scale kinetic-energy budget for 
multiphase flows.

These scale-by-scale studies have revealed cascade mechanisms that are specific to multiphase turbulence. In buoyancy-driven bubbly flows, energy spectra show a dual scaling, with a Kolmogorov-like $k^{-5/3}$ range above the bubble diameter and a pseudo-turbulent $k^{-3}$ range below it \citep{Pandey23PRL}. In isotropic turbulent emulsions, surface tension provides an additional energy-transfer pathway, extracting kinetic energy at large scales through droplet deformation and breakup and returning it at smaller scales through coalescence or interfacial relaxation \citep{Crialesi22JFM,Crialesi23CP,ThiessetVahe25JFM}. Scale-by-scale budgets further show that advective fluxes dominate inertial-range transfer, whereas surface tension and viscosity regulate energy conversion and dissipation near droplet and viscous scales.

Whether these cascade features carry over to immiscible RT turbulence remains unclear. Unlike statistically stationary and nearly homogeneous emulsions or bubbly flows, RT turbulence is strongly unsteady, inhomogeneous, and anisotropic, involving simultaneous mixing-layer growth, interfacial deformation, droplet and bubble formation, and gravitational energy release. Addressing how energy is redistributed among these processes through scale-by-scale density and kinetic-energy budgets is therefore a central objective of the present work.

Beyond interfacial modulation of multiphase energy cascade, the breakup and coalescence of individual bubbles/droplets and swarms \citep{Vela22SciAdv}, their size distribution \citep{Skartlien13JCP}, and their effects on turbulent statistics \citep{DoddFerrante16JFM} and rheological properties \citep{Yi23PTRSA} are crucial to the physics of multiphase flows. They underpin applications such as spray atomization, emulsion rheology, precipitation prediction, and climate modeling, as well as for large-eddy simulations with limited inertial range resolution \citep{Mcclements10AnnRevFood,Ni24ARFM}.
In canonical isotropic turbulence, droplet size distributions follow Hinze-scale-based power laws in which super-Hinze droplets undergo local fragmentation with a $d^{-10/3}$ scaling, while sub-Hinze structures experience nonlocal coalescence with $d^{-3/2}$ scaling \citep{Crialesi22JFM,CannonRosti24JFM}. Beyond homogeneous, isotropic cases, investigations of immiscible two-fluid turbulence in complex configurations are emerging. \cite{Scarbolo15POF} identified a Weber-number threshold governing stable coalescence versus dynamic breakup-coalescence equilibrium in turbulent channel flow. \cite{Yi21JFM} experimentally linked droplet statistics to effective viscosity in high-volume-fraction Taylor-Couette flows. In turbulent shear flow, \cite{Rosti19JFM} validated the Hinze criterion in turbulent shear flows. \cite{Trummler22POF} distinguished gravitational segregation from coalescence through timescale analysis. Despite these advances, droplet statistics and their turbulence modulation in buoyancy-driven flows such as immiscible RT turbulence remain largely unexplored.

The remainder of this paper is organized as follows. Section \ref{sec:eqs_setups} outlines the governing equations and the numerical framework employed in the simulations. In Section \ref{sec:statistics}, we present the global analysis of the flow, including flow visualizations, the evolution of characteristic length scales, and the similarity of global energy budgets with respect to surface tension. Section \ref{sec:scale_decomp} focuses on scale-by-scale analysis. We present coarse-grained budgets to characterize the cascades of scalar variance and kinetic energy, specifically identifying the role of surface tension power in coupling these two quantities. This section also establishes the theoretical link between surface tension power and interface stretch rate. In Section \ref{sec:bub_drop_stat}, we examine the statistics of discrete bubbles and droplets, including their geometry, morphology, and dynamics, and provide direct validation of the relationship between surface tension power and the scalar variance cascade. Finally, we summarize our main findings in Section \ref{sec:conclude}.

\section{Preliminaries and Simulations} \label{sec:eqs_setups}
\subsection{Governing equations and numerical methods} \label{sec:governing_eq}

We model the immiscible two-phase turbulent flow using a single-fluid incompressible formulation. The governing equations are the continuity equation,
\begin{align}
    \nabla\cdot\bu = 0, \label{eq:incompr}
\end{align}
and the momentum equation,
\begin{align}
    \rho\left(\frac{\partial \bu}{\partial t} + \bu\cdot \nabla \bu\right) &= -\nabla P + \nabla\cdot \boldsymbol{\tau}^u + \boldsymbol{f}^\sigma + \rho \boldsymbol{g}, \label{eq:momentum}
\end{align}
where $\bu$, $P$, and $\rho$ are the velocity, pressure, and density fields. The terms on the right-hand side of Eq. \eqref{eq:momentum} are the pressure gradient, the viscous stress tensor $\boldsymbol{\tau}^u = \mu(\nabla\bu + \nabla \bu^T)$, the surface tension force $\boldsymbol{f}^\sigma$, and the gravitational force, with $\mu$ and $\boldsymbol{g}$ being the dynamic viscosity and gravitational acceleration.

The fluid-fluid interface is tracked by the volume-of-fluid (VOF) method, which advects the volume fraction of the dispersed phase (heavy fluid), $c$:
\begin{align}
    \frac{\partial c}{\partial t} + \bu\cdot \nabla c = 0. \label{eq:marker_eq}
\end{align}
The surface tension force is implemented using the continuous surface force (CSF) model, $\boldsymbol{f}^\sigma = \sigma \kappa \nabla c$, where $\sigma$ is the surface tension coefficient and $\kappa$ is the interface curvature \citep{Brackbill92JCP}. The local fluid properties are determined by a linear mixture rule based on $c$:
\begin{align}
    \rho(c) = c\rho_h + (1-c)\rho_l, \quad \text{and} \quad \mu(c) = c\mu_h + (1-c)\mu_l,
\end{align}
where the subscripts $l$ and $h$ denote the light and heavy phases, respectively.

Simulations are performed using the open-source code PARIS-Simulator \citep{Aniszewski21CPC}. This code has been extensively validated and applied to a variety of two-phase turbulent flows, including mixing layers \citep{Jiang21JFM}, jet atomization \citep{Crialesi22IJMF}, and emulsion segregation \citep{Trummler22POF}.
The numerical scheme employs a finite-volume method on a staggered grid, where velocities are located at cell faces and scalar quantities (pressure, volume fraction) at cell centers. Time integration is performed with a second-order predictor-corrector scheme. For spatial discretization, viscous terms are calculated using a second-order central difference scheme, and convective fluxes for momentum are handled with the Superbee flux limiter. Incompressibility is enforced via the Chorin projection method, which requires solving a Poisson equation for pressure using the PFMG multigrid solver in the HYPRE library. The VOF interface is advected using a Lagrangian-explicit scheme, and its curvature is computed with the height-function method. Further details on the numerical implementation can be found in \cite{Aniszewski21CPC}.

\subsection{Spatial coarse-graining} \label{sec:Coarse-graining}

In this work, we analyze scale interactions in immiscible RT turbulence using spatial coarse-graining, also known as the filtering approach. This is a general framework for decomposing a nonlinear system into components at different scales and studying the interactions between them. In the context of turbulence, coarse-graining provides a natural way to understand inter-scale dynamics \citep{Eyink05}.
The core of the method involves convolving a flow variable $\mathbf{a}(\mathbf{x})$ with a low-pass filter kernel $G_\ell$ of characteristic width $\ell$:
\begin{align}
    \overline{\mathbf{a}}_{\ell}(\mathbf{x}) = \int d^d\mathbf{r} \, G_\ell(\mathbf{r}) \, \mathbf{a}(\mathbf{x}-\mathbf{r}).
    \label{eq:filter_op}
\end{align}
where the kernel is a dilated version of a normalized function $G$, defined as $G_\ell(\mathbf{r}) = \ell^{-d}G(\mathbf{r}/\ell)$ in $d$ dimensions. This operation separates the field into a large-scale component, $\overline{\mathbf{a}}_{\ell}(\mathbf{x})$, and a small-scale component, $\mathbf{a}'(\mathbf{x}) = \mathbf{a}(\mathbf{x}) - \overline{\mathbf{a}}_{\ell}(\mathbf{x})$. Similarly, we define a band-pass filtered quantity that isolates scales within the range $[\ell_1, \ell_2]$ as $\overline{\mathbf{a}}_{[\ell_1,\ell_2]} \equiv \overline{\mathbf{a}}_{\ell_1} - \overline{\mathbf{a}}_{\ell_2}$. 

This filtering framework is versatile, encompassing Fourier analysis and wavelet analysis as special cases with specific choices of kernels \citep{AluieEyink09,ZhaoAluie23PRF}. 
In this study, we use a Gaussian kernel of the form:
\begin{align}
    G_\ell(\mathbf{r}) = \left(\frac{6}{\pi\ell^2}\right)^{3/2} \exp\left(-\frac{6|\mathbf{r}|^2}{\ell^2}\right).
\end{align}
For notational convenience, the subscript $\ell$ on filtered quantities will be omitted hereafter unless required for clarity. 

In simulations with non-periodic boundaries, such as the rigid top and bottom walls in our RT configuration, filtering near the boundaries requires special treatment. We extend the fields beyond the physical domain in a manner consistent with the boundary conditions: velocity is set to zero, scalar fields such as density and volume fraction are extended with zero normal gradient, and pressure is extended according to hydrostatic balance. The Gaussian convolution is then applied to the extended fields, and only the filtered values inside the physical domain are retained.

In variable-density flows, the definition of large-scale kinetic energy is non-trivial. While several options exist \citep{ZhaoAluie18}, we adopt the Favre-filtered kinetic energy, $K(\widetilde{\mathbf{u}}) = \frac{1}{2}\overline{\rho}|\widetilde{\mathbf{u}}|^2$, where $\widetilde{\mathbf{u}} = \overline{\rho\mathbf{u}}/\overline{\rho}$ is the Favre density-weighted filtered velocity.
This choice is motivated by two key factors. First, the Favre formulation ensures that kinetic energy scale transfer is independent of viscous effects, a crucial property for robust scale analysis, particularly in flows with high density contrast \citep{ZhaoAluie18}. Second, it simplifies the filtered momentum equation, making it a standard and practical choice in large-eddy simulations of multiphase flows \citep{Saeedipour21IJMF, Suhas25CEJ}. Although $\widetilde{\mathbf{u}}$ is not divergence-free, the benefits of Favre filtering are compelling. Accordingly, our analysis of inter-scale kinetic energy transfer and its connection to interface dynamics will be based on this framework.

\subsection{Simulation configurations}

We simulate two-phase flow within a rectangular domain under a gravitational field, starting with an unstable density stratification. The top and bottom boundaries are no-slip walls, while the lateral boundaries are periodic. For simplicity, the contact angle at the walls is set to $\pi/2$. Initially, the flow is driven by the RT instability, where potential energy is converted into kinetic energy. This leads to an emulsion-like turbulent stage. Subsequently, the decay of turbulence is initiated when viscous dissipation, primarily from coherent structures interacting with the walls, overwhelms the potential energy release. The turbulent emulsion then gradually segregates into two stably stratified, pure phases. Figure \ref{fig:field_viz} (a)-(c) visualizes the density field at early, intermediate, and late times, clearly illustrating the initial RT instability growth, the formation of the turbulent emulsion, and its final segregation.

\begin{table}
  \begin{center}
\def~{\hphantom{0}}
  \begin{tabular}{lcccccccc}
    Case & $L_x\times L_y\times L_z$ & $N_x\times N_y\times N_z$& $\widehat{T}_\mathrm{max}$ & $\Phi$ & $\sigma$ & $\mathrm{Re}_\lambda$ & $\mathrm{We}_\Delta$ & $\mathrm{We}_L$\\
X1 & $1.6\times 1.6\times 3.2$ & $256\times 256\times 512$ & 14.5 & 0.5 & $5\times 10^{-5}$ & 61 & 0.26 & 49\\
X2 & $1.6\times 1.6\times 3.2$& $256\times 256\times 512$ & 14.5 & 0.5 & $1\times 10^{-4}$ &63 & 0.13 &  26\\
X4 & $1.6\times 1.6\times 3.2$& $256\times 256\times 512$ & 14.5 & 0.5 & $2\times 10^{-4}$ &74 &  0.065 & 24\\ 
X8 & $1.6\times 1.6\times 3.2$& $256\times 256\times 512$ & 14.5 & 0.5 &$4\times 10^{-4}$ & 117 & 0.0325 & 33\\
X10 & $1.6\times 1.6\times 3.2$& $256\times 256\times 512$ & 14.5 & 0.5 & $5\times 10^{-4}$ &122 & 0.026 & 28\\
X20 & $1.6\times 1.6\times 3.2$& $256\times 256\times 512$ & 14.5 & 0.5 & $1\times 10^{-3}$ &171 & 0.013 & 26\\
X30 & $1.6\times 1.6\times 3.2$& $256\times 256\times 512$ & 14.5 & 0.5 & $1.5\times 10^{-3}$ &165 & 0.0087 & 21\\
X100 & $1.6\times 1.6\times 3.2$& $256\times 256\times 512$ & 14.5 & 0.5 & $5\times 10^{-3}$ &354 & 0.0026 & 14\\
X150 & $1.6\times 1.6\times 3.2$& $256\times 256\times 512$ & 14.5 & 0.5 & $7.5\times 10^{-3}$ &1.2 & 0.0017 & $1.9\times 10^{-4}$\\
X1Ratio4 & $1.6\times 1.6\times 6.4$ & $256\times 256\times 1024$ & 27.2  & 0.5 & $5\times 10^{-5}$ & 74 & 0.26 & 69\\
X1$\Upphi 05$ & $1.6\times 1.6\times 3.2$& $256\times 256\times 512$ & 14.5 & 0.05 & $5\times 10^{-5}$ & 33 & 0.26 & 3\\
X1$\Upphi 95$ & $1.6\times 1.6\times 3.2$& $256\times 256\times 512$ & 14.5 & 0.95 & $5\times 10^{-5}$ & 42 & 0.26 & 9\\
  \end{tabular}
  \caption{Parameters of the numerical simulations. For all cases, the densities of the light and the heavy fluids are fixed at $\rho_l=0.9, \rho_h=1.0$, the density difference is $\Delta\rho=\rho_h-\rho_l$, the mean density is $\rho_m=(\rho_l+\rho_h)/2$, and the kinematic viscosity is spatially constant and fixes at $2.55\times 10^{-4}$. The Taylor-scale ($\lambda$, defined in equation~\ref{eq:scales}) Reynolds number is $\mathrm{Re}_\lambda=u'\lambda/\nu$, the grid Weber number is $\mathrm{We}_\Delta = \frac{\rho g L^{1/3}\Delta x^{5/3}}{\sigma} \frac{\Delta\rho}{2\rho_m}$, and the integral scale Weber number is $\mathrm{We}_L = \frac{\rho_l u'^2l}{\sigma}$ where $u'$ is the root-mean-square turbulent velocity and $l$ is the integral length scale. The Reynolds and Weber numbers are evaluated at the instants of maximum dissipation: approximately $\widehat{t}=5.8$-$6.6$ for cases X1-X150, $\widehat{t}=11.8$ for case X1Ratio4, and $\widehat{t}\approx 8$ for cases X1$\Upphi 05$ and X1$\Upphi 95$. The column $\widehat{T}_{\max}$ denotes the maximum nondimensional time reached in each simulation. The column $\Upphi$ denotes the volume fraction of the heavy fluid.}
  \label{tab:parameter}
  \end{center}
\end{table}

The parameters for the numerical simulations presented in this paper are summarized in Table \ref{tab:parameter}. We performed two distinct sets of simulations. The first set, X1–X150, consists of binary mixtures with a fixed heavy fluid volume fraction of 0.5, where each case differs only by its surface tension coefficient. The second set, X1$\Upphi 05$ and X1$\Upphi 95$, is designed to study droplet and bubble dynamics at dilute (heavy fluid fraction $\Upphi=0.05$) and dense ($\Upphi=0.95$) volume fractions of the heavy fluid.
In all simulations, the initial fluid interface was perturbed with random-phase sinusoids over a wavenumber range of $k \in [15, 30]$, with modal amplitudes scaling as $k^{-1}$ \citep{Alphagroup}. The grid resolution is confirmed to be adequate, as indicated by a grid Grashof number of $Gr = \Delta\rho g\Delta^3 / (\rho_m \nu^2) = 0.39 < 1$, which is sufficient for resolving buoyancy-induced turbulence. For each simulation, the Taylor-scale Reynolds number ($\mathrm{Re}_\lambda$) and the integral-scale Weber number ($\mathrm{We}_L$) are evaluated at the instant of maximum turbulent dissipation. A more detailed statistical analysis of these cases will be presented in the following sections.

\section{Temporal evolution of immiscible RT turbulence} \label{sec:statistics}

\begin{figure}
\centering 
\begin{minipage}[b]{1.0\textwidth}  
\centering
\subfigure{\includegraphics[height=3.5in]{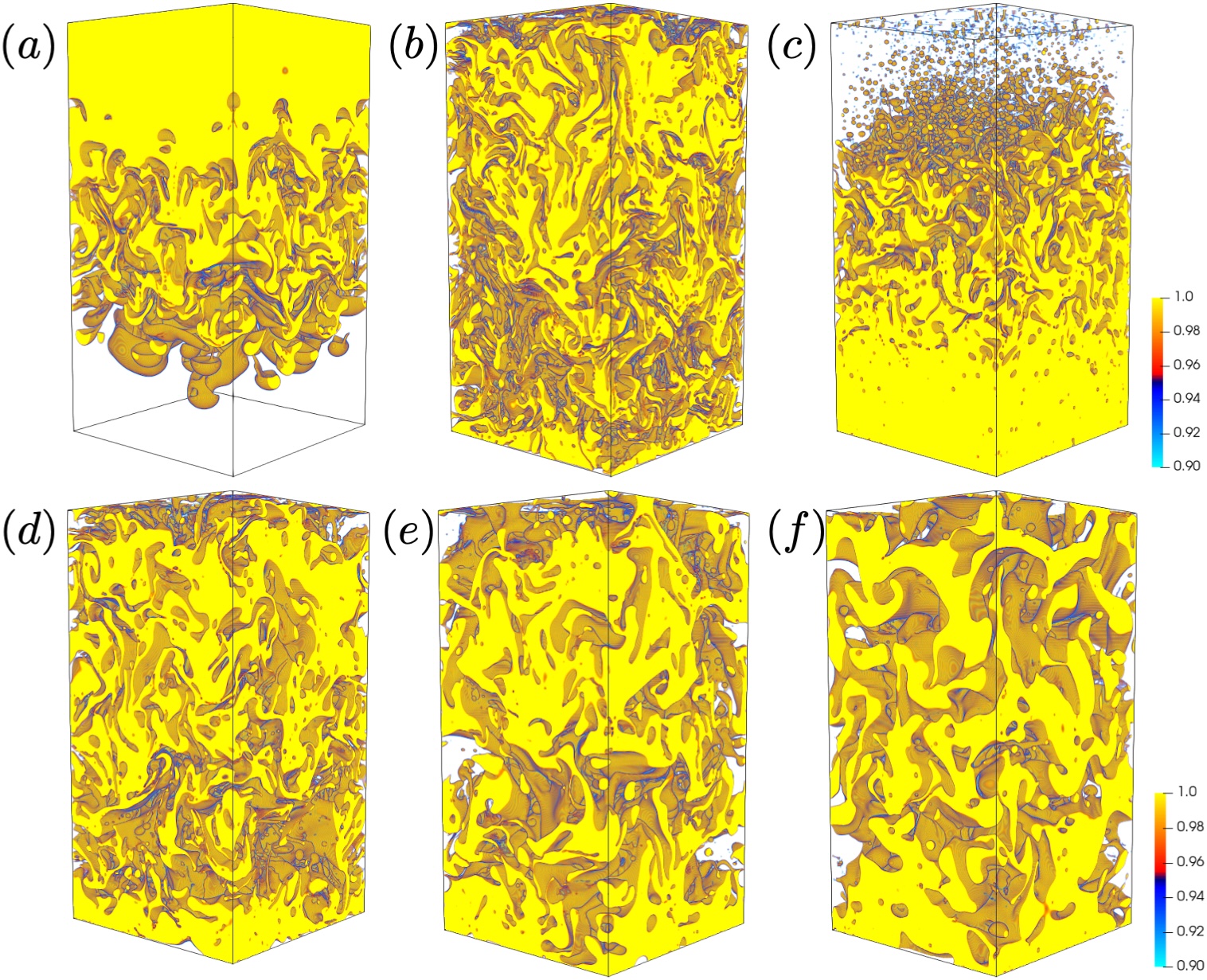} } 
\caption{Visualizations of the density fields. Panels (a)-(c) show the density visualizations for case X1 at non-dimensional time $\widehat{t}=t\sqrt{\frac{A_t g}{L_x}}=2.7,~ 6,~ 13.6$, respectively, with corresponding Taylor-scale Reynolds numbers $\mathrm{Re}_\lambda=17.4, 54.0$, and $17.1$. Panels (d)-(f) show the density visualizations at non-dimensional $\widehat{t}=6$ for cases X2, X4, and X8, respectively, with corresponding $\mathrm{Re}_\lambda=58.2$, $62.1$, and $62.2$. \label{fig:field_viz}}
\end{minipage}
\end{figure}

The simulations listed in Table \ref{tab:parameter} are transient, capturing both the growth and subsequent decay of turbulent kinetic energy. Figure \ref{fig:field_viz} presents representative snapshots of the density field to illustrate this evolution. Panels (a)–(c) depict three distinct stages in simulation case X1: (a) initial RT growth, characterized by classical bubble and spike structures; (b) the turbulent emulsion stage, marked by strong interfacial deformations and the formation of abundant bubbles and droplets; and (c) the late-time decay and phase segregation stage.
These snapshots correspond to non-dimensional times $\widehat{t} = 2.7, 6, 13.6$, respectively. Here, $\widehat{t}\equiv t/\tau$, where $\tau=\sqrt{L_x/(A_t g)}$ represents the characteristic RT time scale, and $A_t=(\rho_h-\rho_l)/(\rho_h+\rho_l)$ is the Atwood number, with $\rho_l, \rho_h$ densities of the light and the heavy fluids.
To demonstrate the effect of surface tension, panels (d)–(f) display snapshots from simulations X2, X4, and X8 at the same non-dimensional time as panel (b). A visual comparison reveals that the characteristic length scale of the flow structures increases with the surface tension coefficient $\sigma$. This indicates that surface tension imposes a small-scale cutoff on density (and volume fraction) fluctuations. Additional visualizations for cases X2, X4, and X8 at $\widehat{t}=13.6$ are provided in figure~\ref{Appfig:late_time_viz} in the Appendix, showing that the flow is close to a stably stratified state at this late time.

\subsection{Evolution of the mixing width}

\begin{figure}
\centering 
\begin{minipage}[b]{1.0\textwidth}  
\centering
\subfigure{\includegraphics[height=1.8in]{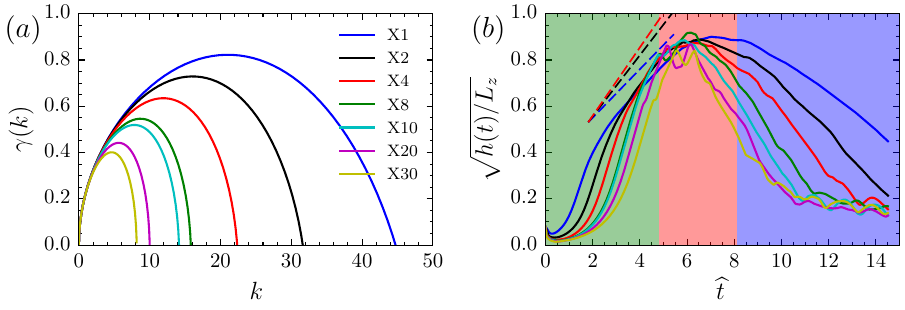}}
\caption{(a) Linear RT growth rate from equation (\ref{eq:growth_rate}) as a function of perturbation wavenumber $k$, including viscous and surface-tension effects. (b) Time evolution of the mixing width $h(t)$; the green, red, and blue shaded intervals denote the RT growth, turbulent-emulsion, and segregation stages, respectively. Dashed lines indicate linear fits to $\sqrt{h(t)}$ over $\widehat{t}\in[2,5.4]$, yielding $\alpha=0.034$, $0.043$, and $0.048$ for cases X1, X2, and X4 (blue, black, and red), respectively. \label{fig:mixing_width}}
\end{minipage}
\end{figure}

During the initial linear stage, the growth rate of a single-mode RT perturbation, including both viscous and surface-tension effects \citep{BellmanPennington54QAM,Sohn09PRE}, is
\begin{align} \label{eq:growth_rate}
\gamma(k)= -\nu k^2 + \sqrt{A_tkg-\frac{\sigma k^3}{\rho_l+\rho_h}+(\nu k^2)^2}.
\end{align}
In this regime, different perturbation modes evolve independently. Figure~\ref{fig:mixing_width}(a) shows according to equation~(\ref{eq:growth_rate}) that, for fixed surface tension, the growth rate $\gamma(k)$ increases with wavenumber at small $k$, then decreases and vanishes at the cutoff wavenumber $k_c=\sqrt{g(\rho_h-\rho_l)/\sigma}$. As $\sigma$ increases, both the overall growth rates and the cutoff wavenumber decrease monotonically, indicating a progressively narrower and slower-growing unstable band. This trend is also evident in the early-time evolution of the mixing width in figure~\ref{fig:mixing_width}(b), where the onset and growth of the mixing zone are increasingly delayed as $\sigma$ increases. 

In particular, from equation~(\ref{eq:growth_rate}), the instability is suppressed in the inviscid limit when $\sigma \geq \sigma_{\rm crit}(k)=\frac{(\rho_h-\rho_l) g}{k^2}$. For the domain-scale mode, $\lambda=L_x=2\pi/k$, this gives 
\begin{align} \label{eq:sigma_crit}
\sigma_{\rm crit}=\frac{(\rho_h-\rho_l) g L_x^2}{4\pi^2}.
\end{align}
As shown in Table~\ref{tab:parameter}, case X30 ($\sigma/\sigma_{\rm crit}=0.23$) remains well below marginal stability, whereas  case X100 ($\sigma/\sigma_\mathrm{crit}=0.77$) lies near the stability threshold and case X150 ($\sigma/\sigma_\mathrm{crit}=1.15$) is fully suppressed. We therefore focus mainly on cases X1-X30 ($\sigma/\sigma_\mathrm{crit}=0.0077\text{-}0.23$), with cases X100 and X150  included for reference.

The linear behavior in the early RT stage persists until the perturbation amplitude reaches roughly one-tenth of the wavelength \citep{Zhou17-1}, at which point nonlinear effects become important. The subsequent dynamics are then dominated by nonlinear interactions, leading to the emergence of coherent rising bubbles and descending spikes, accompanied by an increasing level of multi-scale interactions. The vertical extent of these structures is defined as the mixing width, $h(t)$. At sufficiently high Reynolds numbers and with high-wavenumber initial perturbations, the RT instability reaches a self-similar state where the mixing width grows quadratically: $h(t)= \alpha A_t g t^2$. Here, $\alpha$ is a growth coefficient. For 3-D miscible RT instability, $\alpha$ typically ranges from 0.02 to 0.04 \citep{Alphagroup}, whereas for immiscible fluids, it increases to approximately $0.04 \text{--}0.05$ \citep{YoungHam06JOT}.

For the mixing width, we adopt an integral definition that remains valid throughout the simulation, encompassing both the growth and decay phases \citep{CabotCook06NatPhy}:
\begin{align}
     h(t) = 2\int_{-\infty}^\infty \mathrm{min}(\langle c\rangle_{xy}(z,t), 1-\langle c\rangle_{xy}(z,t))dz
\end{align}
where $\langle c\rangle_{xy}(z,t)$ denotes the horizontal average of the volume fraction field. 
Figure \ref{fig:mixing_width}(b) shows the time evolution of the mixing width for cases X1–X30, which all display the same overall trend. At early times, $h(t)$ increases exponentially, followed, in the low surface tension cases (X1–X4 in dashed lines), by a quadratic growth interval. As the mixing layer thickens, wall confinement increasingly inhibits further expansion and $h$ saturates at a maximum of about $0.8$. Beyond this peak, RT-driven turbulence decays and the emulsion segregates, so $h(t)$ decreases toward zero as the flow relaxes to a separated, stably stratified configuration. Consistent with this progression, we view the evolution in figure \ref{fig:mixing_width}(b) as moving from an RT growth regime (shaded green) to a turbulent emulsion stage (red), and finally to a segregation regime (blue) marked by decaying turbulence and phase separation.

The dashed lines in figure \ref{fig:mixing_width}(b) highlights the early-time evolution of the mixing width. The low–surface-tension cases (X1–X4) show a distinct quadratic-growth interval, whereas the higher–surface-tension cases (X8–X30) do not, because larger $\sigma$ confines the unstable spectrum to long wavelengths (small $k$, see figure \ref{fig:mixing_width}(a)) and, at the moderate Reynolds numbers considered here, prevents the emergence of the self-similar scaling observed at lower surface tension \citep{YoungHam06JOT}. For X1–X4, $\sqrt{h(t)}$ grows approximately linearly over a finite window, with $\alpha=0.034$, $0.043$, and $0.048$ for X1, X2, and X4, respectively from numerical fits. These coefficients exceed values typically reported for miscible RT flows, consistent with surface tension suppressing small-scale mixing and preserving coherent bubble structures \citep{Briard2024JFM}. Moreover, $\alpha$ increases with $\sigma$, since higher surface tension shifts the dominant mode to longer wavelengths and the bubble/spike terminal velocity increases with wavelength in two-phase RT flows \citep{Sohn09PRE}. As $\sigma$ increases, this quadratic-growth window also narrows and vanishes entirely for X8 and above.

\subsection{Characteristic length scales and interfacial area}
In the evolution of two-phase turbulence, multi-scale interactions lead to an extended inertial range and the emergence of characteristic length scales with distinct scale separation.
The key characteristic length scales in the two-phase turbulent systems are the Kolmogorov scale ($\eta$), the Taylor microscale ($\lambda$), and the Hinze scale ($l_H$). In the current context, these scales are defined as:
\begin{align} \label{eq:scales}
\eta = (\rho_l \nu^3 / D)^{1/4}, \quad \lambda = \left[ \frac{\langle u_z'^2\rangle}{\langle (\partial u_z'/\partial z)^2 \rangle}\right]^{1/2}, \quad l_H = 0.725(\sigma/\rho_l)^{3/5} (D/\rho_l)^{-2/5}
\end{align}
where $D$ is the mean kinetic energy dissipation rate, $u_z'$ is the root-mean-square vertical velocity fluctuation, $\nu$ is the kinematic viscosity, and $\rho_l$ is the light fluid density. Here, the Taylor scale is evaluated from the vertical velocity gradient, while the horizontal Taylor scales are smaller than the vertical one due to flow anisotropy \citep{ZhouCabot19POF}. Previous studies on miscible RT turbulence \citep{ZhaoAluie22JFM,ZhaoLi25JFM} have confirmed that the classical definitions for the Kolmogorov and Taylor scales remain applicable. In this work, we will demonstrate that, in addition to $\eta$ and $\lambda$, the above Hinze scale definition similarly provides a crucial benchmark for immiscible RT turbulence, effectively predicting the scale at which surface tension transitions from acting as a sink to a source of kinetic energy.

\begin{figure}
\centering 
\begin{minipage}[b]{1.0\textwidth}  
\centering
\subfigure{\includegraphics[height=3.6in]{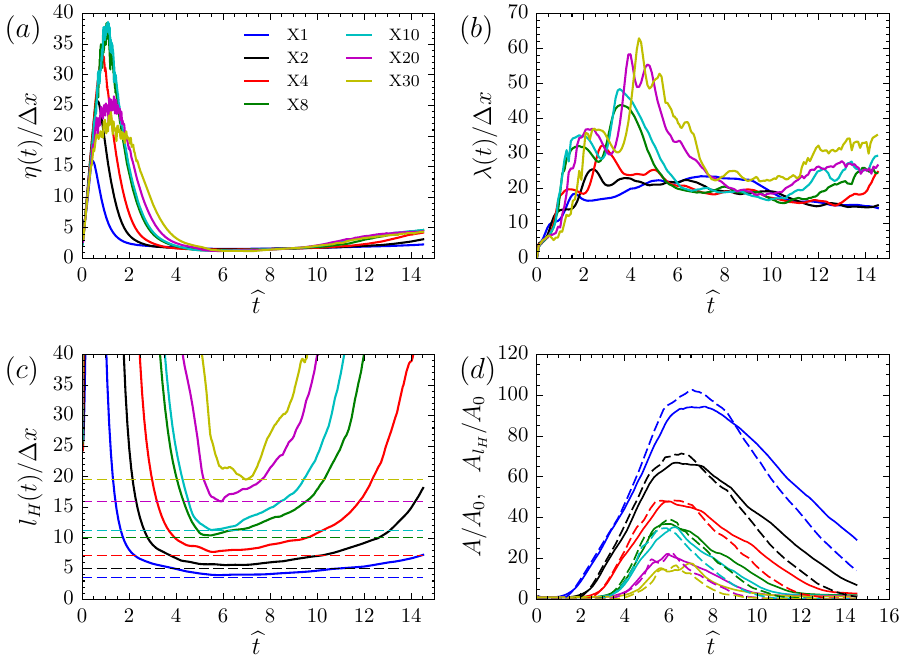} } 
\caption{Temporal evolution of the (a) Kolmogorov scale $\eta$, (b) Taylor microscale $\lambda$, (c) Hinze scale $l_H$, and (d) interfacial area $A$. In panel (c), the horizontal dashed lines corresponds to the capillary scale  $l_c = \sqrt{\frac{\sigma}{(\rho_h-\rho_l)g}}$. In panel (d), solid lines represent areas calculated from simulations X1–X30, while dashed lines indicate the estimated area $\widehat{A}_{l_H}$ in equation~(\ref{eq:area_estimation}) based on the respective Hinze scale. \label{fig:scales}}
\end{minipage}
\end{figure}

Figure \ref{fig:scales} presents the temporal evolution of the characteristic length scales defined in equation~(\ref{eq:scales}), normalized by the grid size $\Delta x$. In panel (a), the normalized Kolmogorov scale satisfies $\eta/\Delta x > 1.5/\pi$ for all simulation cases (X1–X30). This confirms that the velocity fields are well-resolved according to the criterion of \cite{Pope2001}. The evolution of $\eta$ follows a consistent trend across all cases: a sharp rise and subsequent fall during the initial transient, followed by a plateau characteristic of well-developed turbulence ($5<\widehat{t} <10$). Finally, the scale gradually increases during the decaying phase ($\widehat{t} > 10$). Notably, the duration of the well-developed turbulence phase shortens as the surface tension coefficient increases, reflecting the modulation of the flow field by surface tension.

The Taylor microscale $\lambda$ in panel (b) displays a stronger dependence on surface tension than the Kolmogorov scale. Specifically, $\lambda$ increases with $\sigma$, consistent with the physical intuition that the velocity field becomes spatially smoother in high-surface-tension regimes. Low surface tension cases (X1–X4) show a gradual temporal change. In contrast, high surface tension cases (X8–X30) exhibit more drastic changes in time. The Taylor scales in different cases intersect during the turbulent emulsion stage ($\widehat{t} \in [6, 10]$). This crossing likely occurs because the interaction between large-scale RT structures and the domain walls begins to dominate over the effects of surface tension modulation.

Panel (c) of figure~\ref{fig:scales} shows that the Hinze scale $l_H$ increases with surface tension $\sigma$, consistent with equation~(\ref{eq:scales}). In the high-$\sigma$ cases, $l_H$ rises more rapidly after the plateau, indicating a faster decay of turbulent fluctuations. For reference, the capillary length $l_c \equiv \sqrt{\sigma/(\Delta \rho\, g)}$ is shown as dashed lines. Notably, the minimum Hinze scale reached during the turbulent-emulsion stage is close to $l_c$, suggesting that breakup is ultimately limited by capillary stabilization under gravity-driven forcing. In immiscible RT turbulence, energy is injected by buoyancy and transferred through an approximately Kolmogorov cascade \citep{Chertkov05PRE}. During the growth phase the dissipation increases, driving $l_H$ downward, but surface tension progressively suppresses fragmentation at small scales. Because the large-scale driving accelerations remain $O(g)$, the cascade cannot sustain stable droplets much smaller than the scale at which surface tension balances gravity, i.e., $l_c$. Consequently, $\min(l_H)$ approaches $l_c$ near peak turbulent intensity. This trend persists in the extended-domain simulation X1Ratio4 (figure~\ref{Appfig:Hinze_ratio4} in the Appendix), supporting $l_c$ as a robust lower bound for $l_H$ in this regime.

In addition to characteristic length scales, the total interfacial area, $A(t)$, is a key quantity for characterizing RT turbulence and mixing. Figure \ref{fig:scales}(d) displays the evolution of $A(t)/A_0$ (solid lines), where $A(t)$ is calculated as the spatial integral of the local surface density field $|\nabla c|$ \citep{Trummler22POF}, and $A_0\equiv L_x L_y$ is the cross-sectional area. Qualitatively similar to the mixing width, the interfacial area increases during the RT growth stage, saturates during the turbulent emulsion stage, and subsequently decreases during the segregation stage. The interfacial area is significantly larger for low surface tension cases. This is attributed to a reduced Hinze scale, which permits the proliferation of small-scale structures in the density field, as visualized in Figure \ref{fig:field_viz}. 

We further estimate the interfacial area by combining the Hinze scale with the mixing width:
 \begin{align} \label{eq:area_estimation}
    A_{l_H} = \frac{h(t)}{l_H} A_0,
\end{align}
 where $A_0 h(t)$ represents the volume of the emulsion within the mixing region.  This model assumes that, within the turbulent emulsion, the two fluids form an interconnected structure of length scale $l_H$, rather than a dilute suspension of isolated droplets. The dashed lines in figure~\ref{fig:scales}(d) show the estimation of the interfacial area agrees qualitatively with the simulations, particularly during the nonlinear RT growth stage. Noticeable deviations arise during the turbulent emulsion and subsequent segregation stages. As illustrated in figure~\ref{fig:field_viz}(b)–(f), isolated droplets appear at these later times, violating the space-filling assumption underlying equation~(\ref{eq:area_estimation}).

\subsection{Bulk kinetic energy budget}
From equations (\ref{eq:incompr}) and (\ref{eq:momentum}), we derive the kinetic energy budget equation:
 \begin{align} \label{eq:KE_budgets}
     \partial_t \left(\frac{1}{2}\rho|\bu|^2\right) + \nabla\cdot \left(\frac{1}{2}\rho \bu |\bu|^2 + P \bu- \boldsymbol{\tau}^u \cdot\bu\right) = -D + \psi_\sigma + \epsilon^\mathrm{inj}
 \end{align}
The terms on the right-hand side represent the viscous dissipation rate $D=\boldsymbol{\tau}^u:\boldsymbol{S}$, the rate of work done by surface tension $\psi_\sigma=\bu\cdot \boldsymbol{f}^\sigma$ (hereafter `surface tension power' for brevity), and the energy injection rate due to buoyancy $\epsilon^\mathrm{inj} = \rho \bu\cdot \boldsymbol{g}$. Averaging these terms over the domain volume $V$ and time $t$ yields the change in internal energy $\Delta I=\int_0^t \langle D\rangle \mathrm{d}t$, the change in surface energy $\Delta \Psi=-\int_0^t \langle\psi_\sigma\rangle \mathrm{d}t$, and the released potential energy $\Delta P=\int_0^t\langle\epsilon^\mathrm{inj}\rangle \mathrm{d}t$. Similarly, the change in total kinetic energy is $\Delta K=\int_0^t\langle \frac{\partial}{\partial t}\left(\frac{1}{2}\rho|\bu|^2\right)\rangle \mathrm{d}t$. Here, $\langle \cdot \rangle$ denotes the domain average, defined as $\langle \cdot \rangle = V^{-1}\int \cdot \, dV$. Applying this average and integrating equation~(\ref{eq:KE_budgets}) in time gives
\begin{equation}
    \Delta K = -\Delta I - \Delta \Psi + \Delta P .
\end{equation}

\begin{figure}
\centering 
\begin{minipage}[b]{1.\textwidth}  
\centering
\subfigure{\includegraphics[height=3.6in]{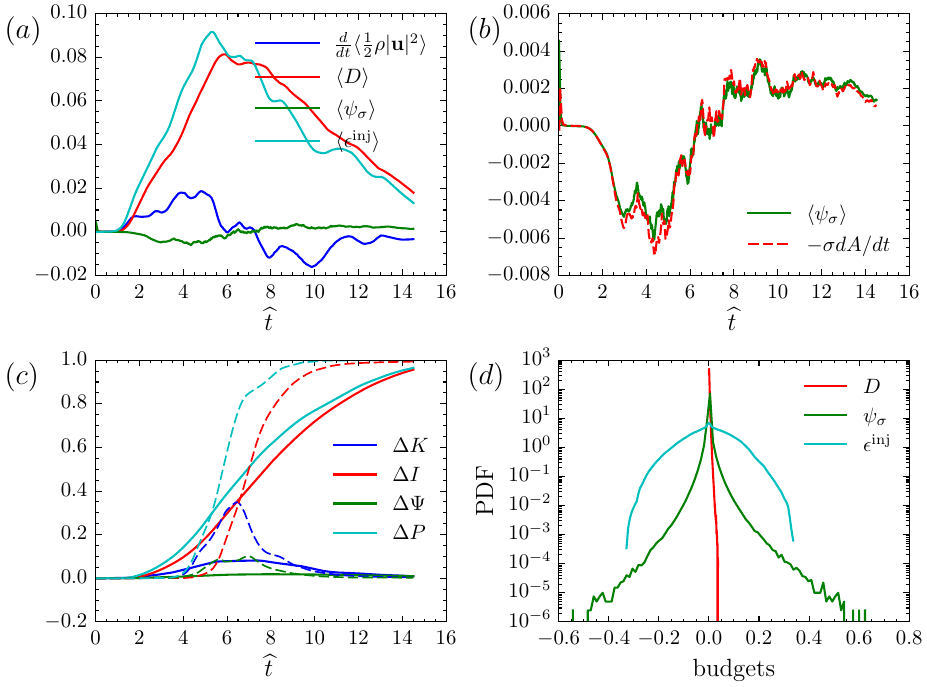} } 
\caption{Kinetic-energy budgets for case X1. (a) Mean instantaneous budget terms from equation~(\ref{eq:KE_budgets}), normalized by $\varepsilon_0=\rho_l u_\mathrm{RT}^3/L_z$, with $u_\mathrm{RT}=\sqrt{A_t g L_z}$. (b) Normalized mean surface-tension power and time derivative of the total interfacial area multiplied by $\sigma$. (c) Cumulative energy balance, normalized by the total potential-energy release; solid and dashed lines denote cases X1 and X30, respectively. (d) PDFs of the three right-hand-side terms in equation~(\ref{eq:KE_budgets}) at $\widehat{t}=6$. \label{fig:bulk_KE_budgets}}
\end{minipage}
\end{figure}

Figures~\ref{fig:bulk_KE_budgets}(a) and (c) show the temporal evolution of the mean instantaneous and cumulative kinetic-energy budgets for case X1, respectively. Panel (c) additionally includes case X30 in dashed lines, which exhibits trends similar to those of case X1 but over a shorter duration. The instantaneous budget is normalized by $\varepsilon_0=\rho_l u_\mathrm{RT}^3/L_z$, where $u_\mathrm{RT}=\sqrt{A_t g L_z}$ is the characteristic RT velocity scale, while the cumulative budget is normalized by the total potential-energy release. In panel (a), the instantaneous kinetic-energy rate $\frac{d}{dt}\langle \tfrac{1}{2}\rho|\mathbf{u}|^2\rangle$ is positive during the RT growth stage, crosses zero in the turbulent emulsion stage, and becomes negative during the segregation stage as turbulence decays.
Both energy injection and dissipation remain positive throughout the evolution. The injection rate peaks earlier than the dissipation rate, and its magnitude exceeds dissipation during the growth stage but falls below it during the decay stage, reflecting their causal relationship. The work done by surface tension is negative during the growth stage, converting kinetic energy into surface energy by increasing the total interfacial area. In contrast, during the segregation stage, this contribution becomes positive, indicating the release of surface energy back into kinetic energy as the deformed interface relaxes. A magnified view of the surface-tension power is compared with the time derivative of the total interfacial area in figure~\ref{fig:bulk_KE_budgets}(b), confirming the relation $\langle \mathbf{f}^\sigma\cdot \mathbf{u}\rangle = -\sigma \frac{dA}{dt}$. This agreement shows that the rate of change of surface energy is directly proportional to the rate of change of total interfacial area \citep{DoddFerrante16JFM}.

In panel (c), the cumulative kinetic energy $\Delta K$ peaks during the turbulent emulsion stage and decays to nearly zero by the end of the segregation stage. The change in surface energy follows a similar temporal evolution (negative time integral of $\langle\psi_\sigma\rangle$), but with a much smaller magnitude in case X1 and relatively larger in case X30. The released potential energy and the increase in internal energy exhibit comparable evolutions and identical terminal values, confirming that all released potential energy is ultimately dissipated. Although the mean surface energy of the bulk flow is small compared to the other budget terms in case X1, its spatial distribution, shown by the probability density function (PDF) in figure~\ref{fig:bulk_KE_budgets}(d), reveals that its local magnitude exceeds that of the other contributions. This suggests that surface tension plays an important role in high-order statistics, such as vorticity and strain rate. We will show that the scale-by-scale contribution of surface tension is also significant.

\begin{figure}
\centering 
\begin{minipage}[b]{1.0\textwidth}  
\centering
\subfigure
{\includegraphics[height=3.6in]{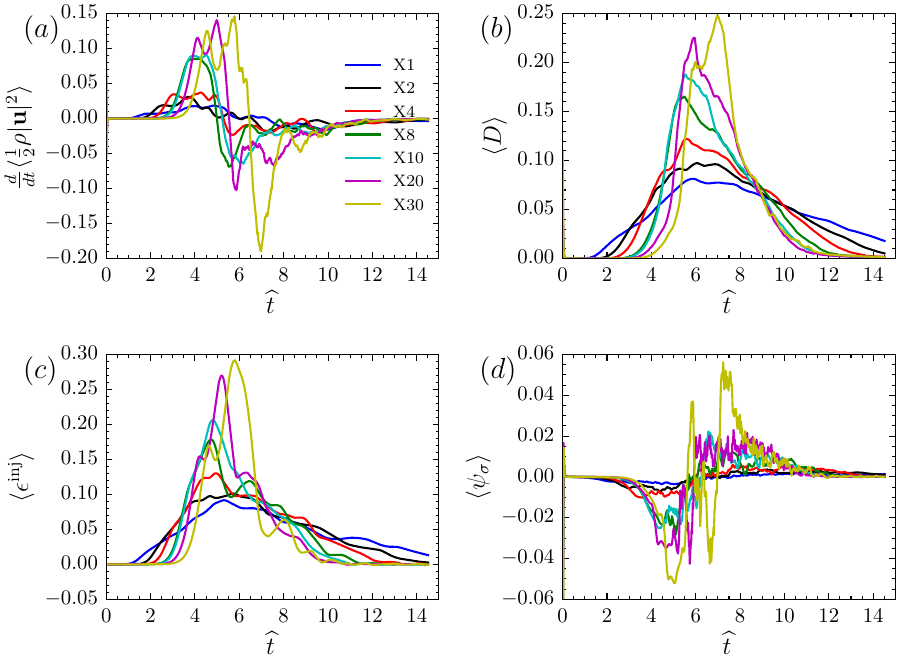} } 
\caption{Temporal evolution of mean instantaneous budgets for simulations X1–X30: (a) Time derivative of KE; (b) dissipation rate; (c) energy injection rate; and (d) surface tension power. Normalization is identical to figure \ref{fig:bulk_KE_budgets}(a). \label{fig:ins_budgets}}
\end{minipage}
\end{figure}

\begin{figure}
\centering 
\begin{minipage}[b]{1.0\textwidth}  
\centering
\subfigure
{\includegraphics[height=3.6in]{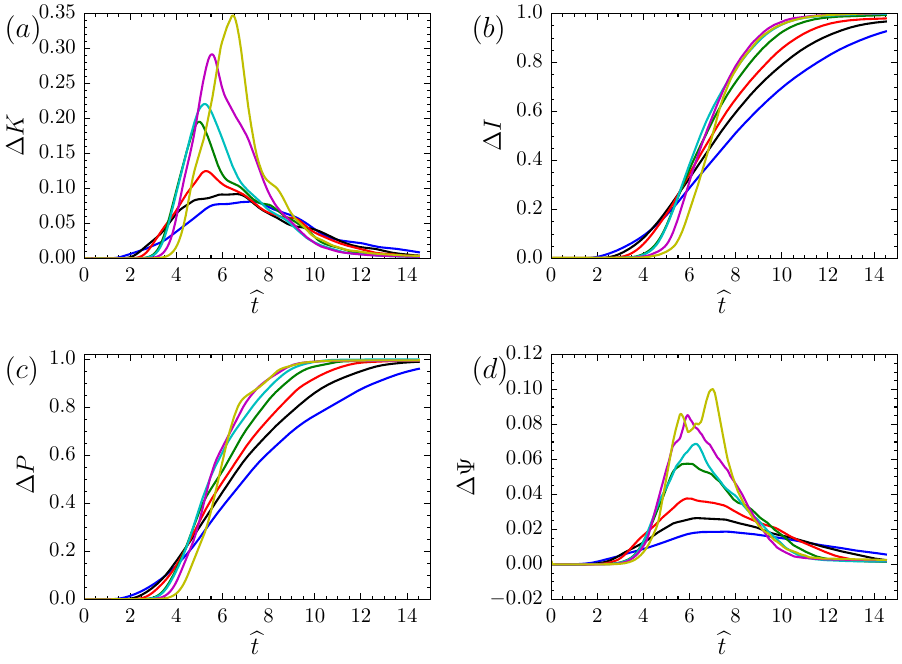} } 
\caption{Temporal evolution of cumulative budgets for simulations X1–X30: increments of (a) kinetic energy; (b) internal energy; (c) potential energy; and (d) excess surface energy. Normalization is identical to figure \ref{fig:bulk_KE_budgets}(c). \label{fig:cumu_budgets}}
\end{minipage}
\end{figure}

Figure~\ref{fig:ins_budgets} shows the effect of surface tension on the instantaneous energy budgets. The rate of mean kinetic-energy change, $\frac{d}{dt}\langle \tfrac{1}{2}\rho|\mathbf{u}|^2\rangle$, mean dissipation rate $\langle D\rangle$, mean energy injection $\langle \epsilon^{\mathrm{inj}}\rangle$, and mean surface-tension power $\langle \boldsymbol{f}^\sigma\cdot\bu\rangle$ all increase in magnitude with increasing surface tension. This trend is consistent with the mixing-width evolution: lower surface tension produces a larger interfacial area, suppressing mixing-layer growth and reducing the mean energy injection rate $\langle \epsilon^{\mathrm{inj}}\rangle=\langle \rho \boldsymbol{g}\cdot\mathbf{u}\rangle$. This leads to lower kinetic energy growth and dissipation rate, so cases with smaller $\sigma$ require longer times to complete segregation.

Figure~\ref{fig:cumu_budgets} presents the cumulative energy budgets, including changes in kinetic, internal, potential, and surface energies. The similar profile shapes across cases, especially in panels (b) and (c), suggest that surface tension sets the characteristic timescale of the bulk energy-budget evolution. Panel (d) further shows that kinetic-to-surface-energy conversion is non-negligible: the excess surface energy $\Delta \Psi$ reaches approximately $0.11$ in case X30, compared with a total potential-energy release of around 1.

\subsection{Scaling of RT statistics with surface tension}
\begin{figure}
\centering 
\begin{minipage}[b]{1.0\textwidth}  
\centering
\subfigure
{\includegraphics[height=3in]{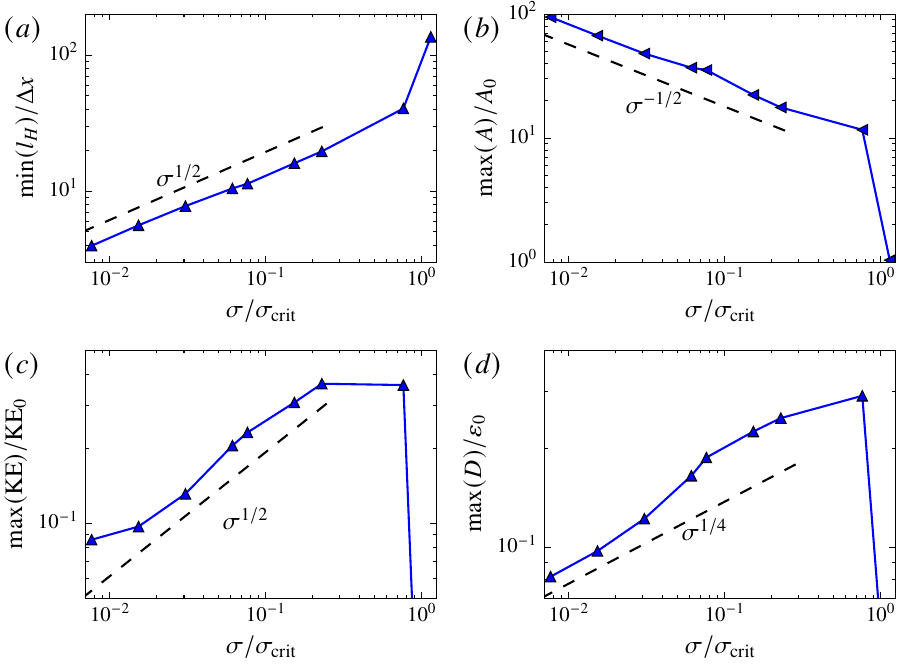} } 
\caption{Scaling of the temporal minimum Hinze scale and the maxima of interfacial area, kinetic energy, and dissipation rate with the surface tension coefficient. The kinetic energy $\mathrm{KE}$ is equal to its increment $\Delta K$, since the flow starts from rest. In panels (c, d), the normalizations are $\mathrm{KE}_0=\frac{1}{2}\rho_l u_\mathrm{RT}^2$ and $\epsilon_0 = \rho_l u_\mathrm{RT}^3/L_z$, respectively. \label{fig:sigma_scaling}}
\end{minipage}
\end{figure}

Figures~\ref{fig:scales}(d) and \ref{fig:ins_budgets}(b,c) show that the interfacial area, mean dissipation rate, and mean energy injection exhibit similar temporal profiles across different surface-tension cases, suggesting that these RT quantities can be collapsed onto a single curve through appropriate renormalization. Figure~\ref{fig:sigma_scaling} shows the minimum Hinze scale, $\min(l_H)$, and the maxima of the interfacial area, $\max(A)$, mean kinetic energy, $\max(\mathrm{KE})$, and mean dissipation rate, $\max(D)$, for cases X1-X150 as functions of $\sigma/\sigma_\mathrm{crit}$, spanning $0.0077$ to $1.15$. Here, $\sigma_\mathrm{crit}$ is the critical surface tension defined in equation~(\ref{eq:sigma_crit}), above which the domain-scale perturbation is suppressed. Within the range $\sigma/\sigma_\mathrm{crit}\in[0.0077,0.23]$, corresponding to cases X1-X30, the numerical results show the scalings $\min(l_H)\propto \sigma^{1/2}$, $\max(A)\propto \sigma^{-1/2}$, $\max(\mathrm{KE})\propto \sigma^{1/2}$, and $\max(D)\propto \sigma^{1/4}$. At larger surface tensions, deviations arise as the flow approaches or exceeds the stability threshold, as in cases X100 and X150.

Assuming that 3-D immiscible RT turbulence satisfies the Kolmogorov energy cascade \citep{Chertkov05PRE}, the inertial range extends down to the Hinze scale, $l_H$, defined in equation (\ref{eq:scales}) as the scale where turbulent dynamic pressure balances surface tension:
 $$ l_H \sim \left(\frac{\sigma}{\rho}\right)^{3/5} D^{-2/5}. $$
At the late-time self-similar stage, the flow at small scales is governed by the balance between gravity and surface tension. Consequently, the minimum characteristic length scale corresponds to the capillary length, $l_c$:
 $$ l_c = \sqrt{\frac{\sigma}{\Delta\rho g}}. $$
By identifying the minimum Hinze scale with the capillary length as observed in figure \ref{fig:scales}(c) (i.e., $\min(l_H) \approx l_c$), we obtain the following balance:
 $$ \left(\frac{\sigma}{\rho}\right)^{3/5} D^{-2/5} \sim \left(\frac{\sigma}{\Delta\rho g}\right)^{1/2}. $$
Solving for the dissipation rate $D$, we derive:
 $$ D \sim C_D \left(\frac{\sigma}{\Delta\rho g}\right)^{-5/4} \left(\frac{\sigma}{\rho}\right)^{3/2} = C_D \Delta\rho^{5/4} g^{5/4} \rho^{-3/2} \sigma^{1/4}.$$
 This yields the scaling relation $D \propto \sigma^{1/4}$, which agrees with our numerical observations. In the above expression, $C_D$ denotes the dimensionless compensated coefficient, defined as
$C_D=\frac{\max(D)\rho_m^{3/2}}
{\Delta\rho^{5/4}g^{5/4}\sigma^{1/4}}$,
where $\rho_m=(\rho_h+\rho_l)/2$ is the reference density. The numerical results show that $C_D$ is approximately 0.4-0.45 for the simulations X1-X30, as is indicated in figure~\ref{Appfig:Cd_sigma} in the Appendix.

Adopting the scaling $D \propto \sigma^{1/4}$, we first consider the Hinze scale, $l_H \sim (\sigma/\rho)^{3/5} D^{-2/5}$, which yields $l_H \propto \sigma^{1/2}$. Similar relation holds for the interfacial area, $A\propto l_H^{-1} \propto \sigma^{-1/2}$ (equation (\ref{eq:area_estimation})).  Next, relating the dissipation to the integral scale via $D \sim u_L^3/L \sim (gL)^{3/2}/L = g^{3/2}L^{1/2}$, we find that $L$ also scales as $\sigma^{1/2}$. Consequently, the kinetic energy, $K \sim (D L)^{2/3}$, follows the scaling $K \sim \sigma^{1/2}$, while the injection term scales as $\epsilon^\mathrm{inj}=\rho u_i g_i\sim \rho g \sqrt{g L}\sim \sigma^{1/4}$. Finally, assuming the total potential energy released during the RT process is constant ($\int D dt = \mathrm{const}$), the duration of the flow scales inversely with the dissipation rate, yielding $T \sim D^{-1} \propto \sigma^{-1/4}$. 
This is consistent with the shorter durations of the RT development and segregation stages observed in high-$\sigma$ cases that remain well below the stability threshold. Thus, the Taylor-scale Reynolds number scales as 
$$R_\lambda \sim \frac{u_{\mathrm{rms}}\lambda}{\nu} \sim \frac{u_{\mathrm{rms}}^2}{\sqrt{\nu D}} \sim \frac{\Delta K}{\sqrt{\nu D}} \sim \sigma^{3/8}.$$

Similarly, the excess surface energy scales as
$\Delta \Psi = \frac{\sigma (A-A_0)}{V} \sim \sigma^{1/2}.$
The surface-tension power can be estimated as $\psi_\sigma = - \frac{d\Delta \Psi}{dt} \sim \frac{\Delta \Psi}{T}$,
where $T$ is a characteristic timescale of the RT evolution. As discussed above, $T\sim D^{-1}\sim \sigma^{-1/4}$. Therefore,
$\psi_\sigma \sim \frac{\Delta \Psi}{T}\sim \sigma^{3/4}.$

\begin{figure}
\centering 
\begin{minipage}[b]{1.0\textwidth}  
\centering
\subfigure
{\includegraphics[height=3in]{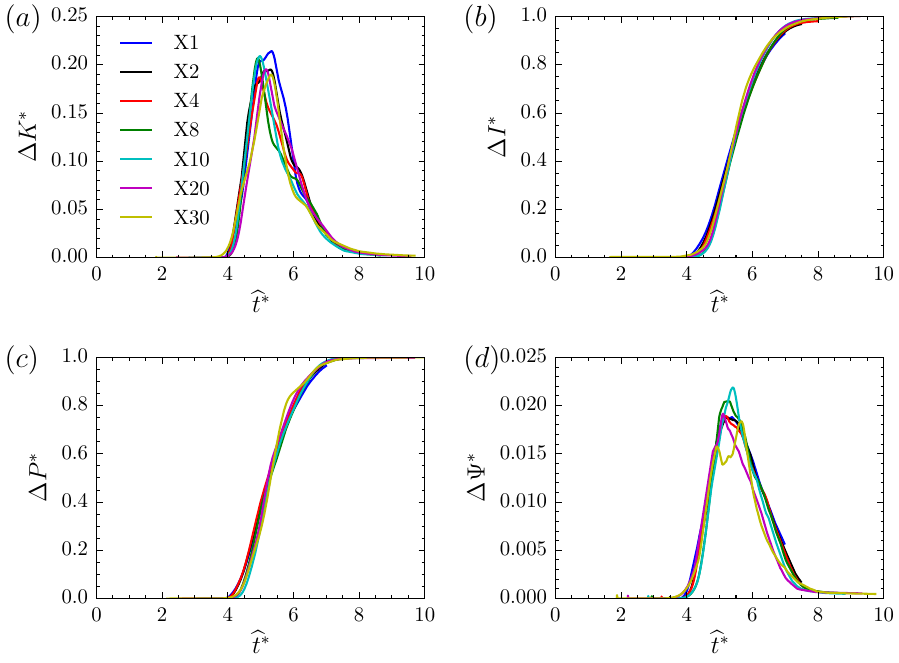} } 
\caption{
Rescaled cumulative kinetic-energy budgets for cases X1–X30. Panels show the contributions from (a) kinetic energy, (b) internal-energy change, (c) potential-energy release, and (d) excess surface energy. Both time and energy are nondimensionalized following equation~(\ref{eq:budget_rescale}). \label{fig:budget_rescaled}}
\end{minipage}
\end{figure}

Based on the derived scaling relations, we rescale the cumulative budgets of cases X1-X30 presented in figure~\ref{fig:cumu_budgets} using the following normalization:
\begin{align}
\begin{split}
    &\widehat{t}^* = (\widehat{t}-\widehat{t}_\mathrm{peak})\mathrm{Bo}^{-1/4} + \widehat{t}_\mathrm{peak}, \\
    &\Delta K^* = \Delta K~ \mathrm{Bo}^{1/2}, \quad
    \Delta I^* = \Delta I, \quad
    \Delta P^* = \Delta P, \quad
    \Delta \Psi^*=\Delta \Psi~ \mathrm{Bo}^{1/2}
\end{split}
\label{eq:budget_rescale}
\end{align}
where $\mathrm{Bo}\equiv \Delta\rho g l_{\mathrm{ref}}^{2}/\sigma$ is the Bond number, with $l_{\mathrm{ref}}$ the characteristic length corresponding to the wavenumber at which the initial perturbation spectrum attains its maximum ($l_{\mathrm{ref}}\approx L_x/24$), and $\widehat{t}_{\mathrm{peak}}$ the time at which the kinetic energy reaches its maximum. Note that $\widehat{t}_\mathrm{peak}$ varies slightly among cases X1–X30, as it depends on both the initial perturbations and the surface tension, and no single universal formula is applicable. In figure~\ref{fig:budget_rescaled}, the rescaled kinetic, internal, potential and surface energies collapse well across different cases.

\subsection{Small scale statistics}

\begin{figure}
\centering 
\begin{minipage}[b]{1.0\textwidth}  
\centering
\subfigure
{\includegraphics[height=1.5in]{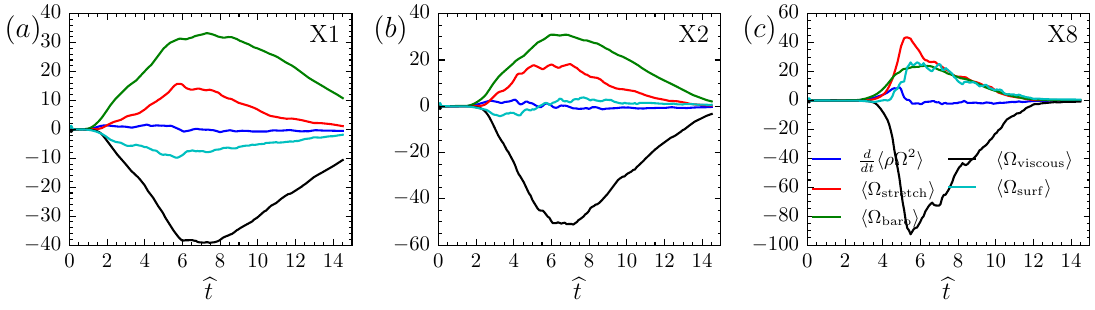} } 
\caption{Mean enstrophy budget evolution for (a) case X1, (b) case X2, and (c) case X8. The budgets are defined in equation~(\ref{eq:ens_budget}). 
\label{fig:vort_budget}}
\end{minipage}
\end{figure}

At small scales, vorticity and the strain-rate tensor govern flow topology and play key roles in the energy cascade and dissipation. Although the budgets of enstrophy and squared strain-rate have been widely studied in canonical flows \citep{Pope2001,Zhao23PoF}, immiscible RT turbulence introduces additional complexity through interfacial forces. To quantify these effects, we examine the evolution equation for enstrophy ($\Omega^2= |\boldsymbol{\omega}|^2/2$) in immiscible RT flows \citep{HasslbergerKlein18JFM}:
\begin{align} \label{eq:ens_budget}
    \rho \frac{D}{Dt}\frac{1}{2}|\bomega|^2 = \underbrace{\rho \bomega\cdot\nabla\bu\cdot\bomega \vphantom{\frac{1}{\rho}}}_{\Omega_\mathrm{stretch}} + \underbrace{\frac{1}{\rho} \bomega\cdot (\nabla\rho\times\nabla P)}_{\Omega_\mathrm{baro}} + \underbrace{\rho\bomega\cdot\nabla\times\left(\frac{1}{\rho}\nabla\cdot\boldsymbol{\tau}^u\right)}_{\Omega_\mathrm{viscous}}+\underbrace{\sigma \boldsymbol{\omega}\cdot \left(\nabla\kappa\times\nabla c\right)\vphantom{\frac{1}{\rho}}}_{\Omega_\mathrm{surf}},
\end{align}
After domain averaging, it reduces to
\begin{equation*}
    \frac{d}{dt}  \left\langle \rho \Omega^2\right\rangle = \langle \Omega_\mathrm{stretch}\rangle +\langle \Omega_\mathrm{baro}\rangle+\langle \Omega_\mathrm{viscous}\rangle+\langle \Omega_\mathrm{surf}\rangle
\end{equation*}
Figure~\ref{fig:vort_budget} shows the temporal evolution of these budget terms for cases X1, X2, and X8. The time derivative of enstrophy reflects the stages of RT mixing: it is positive during the development stage, crosses zero near the turbulent emulsion phase, and becomes negative during phase segregation.

The baroclinic torque ($\Omega_\mathrm{baro}$) and vortex stretching ($\Omega_\mathrm{stretch}$) terms act as positive sources in the spatial mean, while the viscous term ($\Omega_\mathrm{viscous}$) consistently acts as a sink. Comparison of cases X1, X2, and X8 shows that the baroclinic term is relatively insensitive to surface tension, as it is primarily driven by the global density stratification. In contrast, the magnitudes of both vortex stretching and viscous dissipation increase with surface tension. This is because higher surface tension produces larger characteristic structures (larger Hinze scale), which attain higher terminal velocities, leading to a higher effective Reynolds number (Table~\ref{tab:parameter}) and more intensified motion.

The direct contribution of surface tension, $\Omega_\mathrm{surf}$, reveals a competition between interfacial stabilization and capillary-driven vorticity generation. As shown in Figure~\ref{fig:vort_budget}, this term acts primarily as a sink in the low-surface-tension case (X1) but becomes a source at higher surface tension (X2 and X8). This transition corresponds to the changes in flow regimes. During the RT development stage, surface tension primarily suppresses enstrophy growth by resisting interfacial stretching and damping small-scale Kelvin–Helmholtz instabilities. Conversely, during the decay phase, phase segregation dominates through droplet coalescence and interface restoration. These processes rapidly convert stored surface energy into kinetic energy, making surface tension a net enstrophy source. This effect is strongly amplified in high-$\sigma$ cases, where stiffer interfaces drive more violent restoration and sustain high-frequency droplet oscillations.

\section{Multi-scale scalar and kinetic energy transfer} \label{sec:scale_decomp}

The generation and destruction of a wide range of length scales in both the scalar (color function or volume fraction) and the velocity field during immiscible turbulent RT evolution are governed by multi-scale interactions and transfer processes. These processes, which directly shape the spectral distributions of the fields, are strongly modulated by surface tension. In this section, we examine the scale-by-scale budgets of scalar variance and kinetic energy in immiscible RT turbulence, with particular emphasis on the role of surface tension in inter-scale transfer.

\subsection{Scalar and kinetic-energy spectra and structure functions}

\begin{figure}
\centering 
\begin{minipage}[b]{1.0\textwidth}  
\centering
\subfigure
{\includegraphics[height=3in]{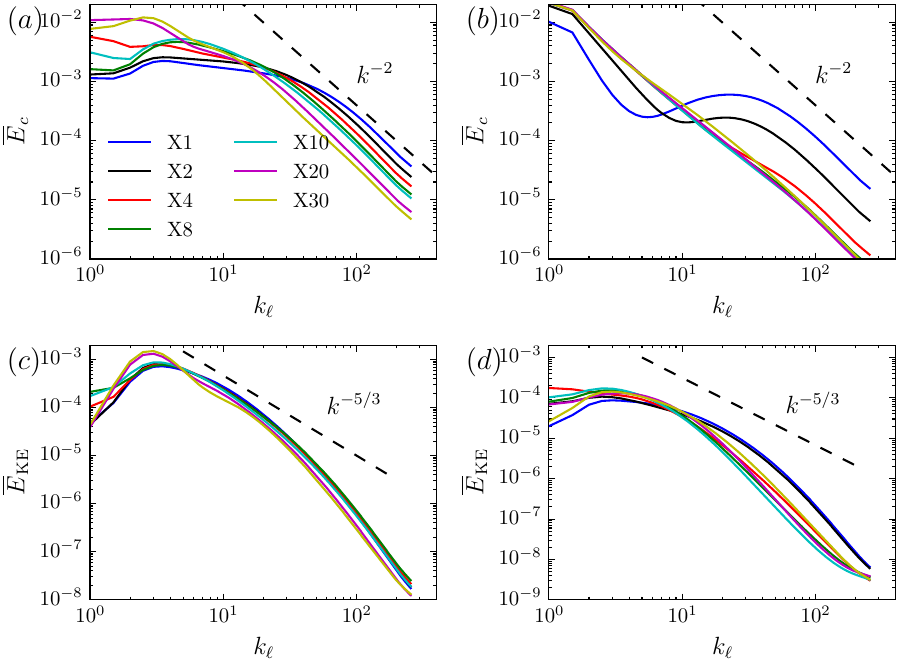} } 
\caption{Filtering spectra of the scalar field (a,b) and the kinetic energy (c,d). Panels (a, c) are calculated at $\widehat{t}=6$, while panels (b,d) are calculated at $\widehat{t}=13.6$. The kinetic energy spectra in (c, d) are normalized to ensure equal total energy across all cases. The filtering wavenumber $k_\ell$ is defined as $k_\ell=L_x/\ell$. The dashed lines indicate reference slopes only; owing to the limited scale separation, the spectra are not intended to establish classical inertial-range scaling \citep{Zhou07PoP}. \label{fig:Ec_Eu_spec}}
\end{minipage}
\end{figure}

 Figure \ref{fig:Ec_Eu_spec} presents the filtering spectra for the scalar and KE fields at two distinct instants, corresponding to the turbulent emulsion stage and the phase segregation stage. The filtering spectra \citep{SadekAluie18PRF} are defined as:
\begin{align}
    \overline{E}_c(k_\ell)=\frac{d}{dk_\ell} \langle |\overline{c}_\ell(\bx)|^2\rangle, \qquad \overline{E}_\mathrm{KE}(k_\ell)=\frac{d}{dk_\ell} \langle \overline{\rho}_\ell |\widetilde{\bu}_\ell(\bx)|^2\rangle/2
\end{align}
where $k_\ell=L/\ell$ is the filtering wavenumber and $\widetilde{\bu}=\overline{\rho\bu}/\overline{\rho}$ represents the Favre density-weighted filtering. 

As shown in figure~\ref{fig:Ec_Eu_spec}(a,b), the scalar spectra scale as $k^{-2}$ at high wavenumbers in all simulation cases. This scaling arises from the near-step function of the density interface inherent in two-phase VOF simulations of immiscible fluids, analogous to the density spectra scaling observed in the presence of strong shocks \citep{WangWan18PRE}. During the emulsion stage ($\widehat{t}=6$, figure~\ref{fig:Ec_Eu_spec}a), the spectra indicate that flows with low surface tension exhibit greater scalar variance at small scales. Conversely, in high surface tension cases, the scalar variance is predominantly contained within the large scales. This is attributed to the stabilizing effect of surface tension, which suppresses small-scale fluctuations. Similarly, during the phase segregation stage ($\widehat{t}=13.6$, figure~\ref{fig:Ec_Eu_spec}b), cases X4–X30 exhibit a $k^{-2}$ scaling across the entire wavenumber range, suggesting that phase segregation is nearly complete. However, for cases X1 and X2, deviations from this scaling imply that small-scale structures persist in the scalar field, indicating that additional time is required to achieve full phase segregation.

As for kinetic energy, the normalized velocity spectra in figure~\ref{fig:Ec_Eu_spec}(c,d) reveal distinct dynamics between the two stages. During the emulsion stage ($\widehat{t}=6$, panel c), the spectra exhibit a limited range of $-5/3$ scaling and a pronounced spectral peak near $k_\ell=3$. This peak reflects the energy-injection process, identifying the scale at which potential energy is actively converted into kinetic energy before cascading to smaller scales. In contrast, during the segregation stage ($\widehat{t}=13.6$, panel d), this conversion process has largely ceased, leading to the disappearance of the spectral peak. As a result, kinetic energy decays more rapidly than in the emulsion stage, and the $-5/3$ scaling is no longer evident. Among the different surface tension cases, X1 and X2 retain relatively higher energy at small scales, which is consistent with the persistence of fine-scale structures in the scalar field shown in figure~\ref{fig:Ec_Eu_spec}(b). Note that due to limited scale separation, the $-5/3$  scaling shown here is not intended to establish classical inertial-range scaling \citep{Zhou07PoP}.

The spectra of the volume-fraction field are closely linked to interfacial geometry. For a sharp two-phase indicator field, the structure function $S_2^c(r)=\left\langle [c(\boldsymbol{x}+\boldsymbol{r})-c(\boldsymbol{x})]^2\right\rangle$ 
measures the probability that two points separated by $r\equiv |\boldsymbol{r}|$ lie in different phases. For a smooth interface, this probability scales with the volume of an $O(r)$ neighbourhood of the interface, giving $S_2^c(r)\sim \frac{A}{V}r$, where $A$ is the interfacial area and $V$ is the domain volume. This leads to the Porod-type scaling \citep{FeiginSvergun1987}
\begin{equation}
\overline E_c(k_\ell)
=\frac{d}{dk_\ell}\left\langle |\overline c_\ell|^2\right\rangle
\sim C A k_\ell^{-2},
\end{equation}
where the coefficient $C$ depends on the filter and normalization. More generally, if the interface is a surface-fractal with dimension $D_s$, then
\begin{equation}
S_2^c(r)\sim r^{3-D_s},\qquad 
\overline{E}_c(k_\ell)\sim k_\ell^{D_s-4}.
\end{equation}

\begin{figure}
\centering 
\begin{minipage}[b]{1.0\textwidth}  
\centering
\subfigure
{\includegraphics[height=1.8in]{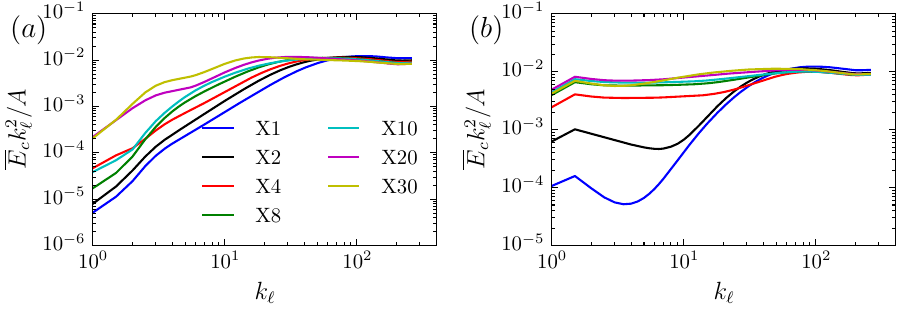} } 
\caption{Compensated scalar filtering spectra, $\overline{E}_c k_\ell^2/A$, at nondimensional times (a) $\widehat{t}=6$ and (b) $\widehat{t}=13.6$, where $A$ is the interfacial area of the corresponding case. \label{fig:Ec_prefactor}}
\end{minipage}
\end{figure}

The observed $k_\ell^{-2}$ scaling at large $k_\ell$ is therefore interpreted as a Porod-type regime, in which the small-scale variance of the volume-fraction field is primarily controlled by interfacial area. To test this interpretation, we examine the compensated spectra $\overline E_c k_\ell^2/A$. As shown in figure~\ref{fig:Ec_prefactor}, these curves approach an approximately case-independent plateau at large $k_\ell$, with values close to $0.01$ at both times. This collapse supports the robustness of the Porod-law prefactor and the use of $\overline E_c$ as a spectral measure of the microstructural content of the dense phase.

\begin{figure}
\centering 
\begin{minipage}[b]{1.0\textwidth}  
\centering
\subfigure
{\includegraphics[height=1.8in]{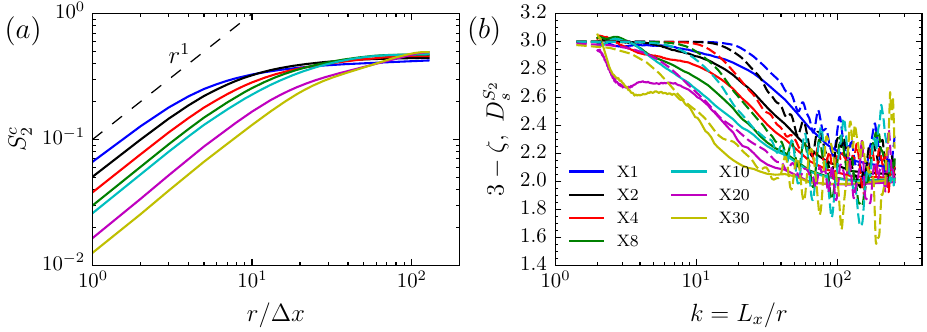} } 
\caption{(a) Structure function of the volume-fraction field at $\widehat{t}=6$ versus increment $r$. (b) Scale-dependent interfacial dimension inferred from $S_2^c(r)$, $D_s^{S_2}=3-\zeta\equiv 3-d\log S_2^c/d\log r$, plotted as a function of $k=L_x/r$ (solid lines), compared with the box-counting estimate $D_s$ for the $c=0.5$ isosurface using boxes of size $r$ (dashed lines).  \label{fig:struct_fractal}}
\end{minipage}
\end{figure}

At intermediate wavenumbers, particularly for the larger Weber-number cases, the local spectral slope varies with scale and does not exhibit an extended plateau. We therefore do not infer a unique fractal dimension directly from $\overline E_c$. Instead, we estimate a scale-dependent interfacial dimension from the structure function,
$$D_s^{S_2}(r)=3-\frac{d\log S_2^c}{d\log r},$$
and compare it with a box-counting estimate of the $c=0.5$ isosurface \citep{ZhaoLi25JFM}. As shown in figure~\ref{fig:struct_fractal}(a), the structure function $S_2^c(r)$ approaches an approximately linear scaling at small separations, consistent with a smooth-interface Porod regime. Over intermediate-to-small scales shown in figure~\ref{fig:struct_fractal}(b), the dimensions inferred from $S_2^c(r)$ and from box counting agree reasonably well, confirming the expected connection between the structure function and interfacial geometry.

\subsection{Coarse-grained equations}

We adopt scale decomposition to investigate the multi-scale dynamics of immiscible RT turbulence. Applying coarse-graining to the scalar field yields the budget equation for the large-scale volume fraction variance:
\begin{align} \label{eq:filtered_csq_budgets}
    \partial_t \left( \frac{1}{2}|\overline{c}_\ell|^2 \right) + \nabla\cdot \left(\frac{1}{2}|\overline{c}_\ell|^2\overline{\bu} +\overline{c}_\ell \overline{\boldsymbol{\tau}}_\ell(c,\bu) \right)= -\Theta_\ell
\end{align}
where $\Theta_\ell(\bx, t)=-\nabla\overline{c}_\ell\cdot\overline{\boldsymbol{\tau}}_\ell(c, \bu)$ represents the transfer of scalar variance across scales, with the sub-scale scalar flux defined as $\overline{\boldsymbol{\tau}}_\ell(c, \bu) = \overline{c \bu}-\overline{c}\ \overline{\bu}$. We note that the scale transfer term is identical for both the squared filtered field, $\frac{1}{2}|\overline{c}_\ell|^2$, and the true resolved variance, $\frac{1}{2}(\overline{c}_\ell-\langle \overline{c}_\ell\rangle)^2$. Consequently, we analyze equation (\ref{eq:filtered_csq_budgets}) as the governing equation for the large-scale scalar variance budget. After domain averaging, equation~(\ref{eq:filtered_csq_budgets}) reduces to
\begin{equation}
    \frac{1}{2}\frac{\partial}{\partial t} \langle|\overline c_\ell|^2\rangle = -\langle\Theta_\ell\rangle 
\end{equation}

The budget equation for the Favre density-weighted filtered kinetic energy at scales larger than $\ell$ is given by:
\begin{align} \label{eq:filtered_KE}
    \partial_t \left( \overline{\rho}_\ell \frac{|\widetilde{\bu}_\ell|^2}{2} \right) + \nabla \cdot \boldsymbol{J}_\ell = -\Pi_\ell - \Lambda_\ell - D_\ell + \epsilon_\ell^\mathrm{inj}+\Psi_{\sigma,\ell}
\end{align}
where $\boldsymbol{J}_\ell$ represents the spatial transport of large-scale kinetic energy. Domain averaging of equation~(\ref{eq:filtered_KE}) leads to 
\begin{equation}
    \partial_t \left\langle \frac{1}{2} \overline{\rho}_\ell |\widetilde{\bu}_\ell|^2 \right\rangle  = -\langle\Pi_\ell\rangle - \langle\Lambda_\ell\rangle - \langle D_\ell\rangle + \langle\epsilon_\ell^\mathrm{inj}\rangle+\langle \Psi_{\sigma,\ell} \rangle.
\end{equation}
The individual budget terms are defined as:
\begin{align} \label{eq:filtered_KE_budgets}
    \begin{split}
        \boldsymbol{J}_{\ell}(\bx)&=\overline{\rho}\frac{|\widetilde{\bu}|^2}{2}\widetilde{\bu} + \bar{P}\overline{\bu} + \bar{\rho}\widetilde{\tau}(\bu,\bu)\cdot \widetilde{\bu}-\overline{\boldsymbol{\tau}}^u\cdot\widetilde{\bu} \\
        \Pi_\ell(\bx)&=-\bar{\rho}\nabla\widetilde{\bu}:\widetilde{\tau}(\bu,\bu); \quad \Lambda_\ell(\bx)=\frac{1}{\bar{\rho}}\nabla \bar{P}\cdot\overline{\tau}(\rho,\bu); \quad \epsilon_\ell^\mathrm{inj}(\bx)=\bar{\rho}\widetilde{\bu}\cdot\widetilde{\boldsymbol{g}};\\
        D_\ell(\bx)&= 2\overline{\mu \boldsymbol{S}}: \nabla\widetilde{\bu}; \quad \Psi_{\sigma,\ell}(\bx)=\widetilde{\bu}\cdot\bar{\boldsymbol{f}^\sigma}
    \end{split}
\end{align}
Here, $\widetilde{\boldsymbol{\tau}}(\bu,\bu)=\widetilde{\bu \bu}-\widetilde{\bu}\,\widetilde{\bu}$ denotes the sub-scale stress tensor. The terms $\Pi_\ell$ and $\Lambda_\ell$ govern the energy transfer between large scales ($\gtrsim \ell$) and small scales ($\lesssim \ell$). Specifically, $\Pi_\ell$ represents the deformation work performed by the sub-scale stress against the large-scale strain, while $\Lambda_\ell$ is the baropycnal work done by the sub-scale mass flux against the large-scale pressure gradient. The term $\Psi_{\sigma,\ell}$ quantifies the work done by surface tension on the large-scale kinetic energy. Consequently, its derivative with respect to the filtering wavenumber $k_\ell\equiv L_x/\ell$, given by $d\Psi_{\sigma,\ell}/dk_\ell$, represents the spectral contribution of surface tension to the kinetic energy at a specific scale $\ell$.

\subsection{Transfer of scalar variance across scales} \label{sec:c_transfer}

In immiscible two-phase flows, the absence of molecular mixing between different phases implies that the total squared scalar volume fraction is strictly conserved, i.e., $\langle c^2\rangle(t) \equiv \langle c^2\rangle(0)$. Consequently, the transfer process described in equation~(\ref{eq:filtered_csq_budgets}) redistributes scalar variance conservatively across length scales without changing the total content. Since $\Theta_\ell$ appears as a sink term in the large-scale budget, $\langle\Theta_\ell\rangle > 0$ denotes an average forward transfer of variance (from large to small scales), while $\langle\Theta_\ell\rangle < 0$ represents a net inverse transfer (from small to large scales).

\begin{figure}
\centering 
\begin{minipage}[b]{1.0\textwidth}  
\centering
\subfigure{\includegraphics[height=1.35in]{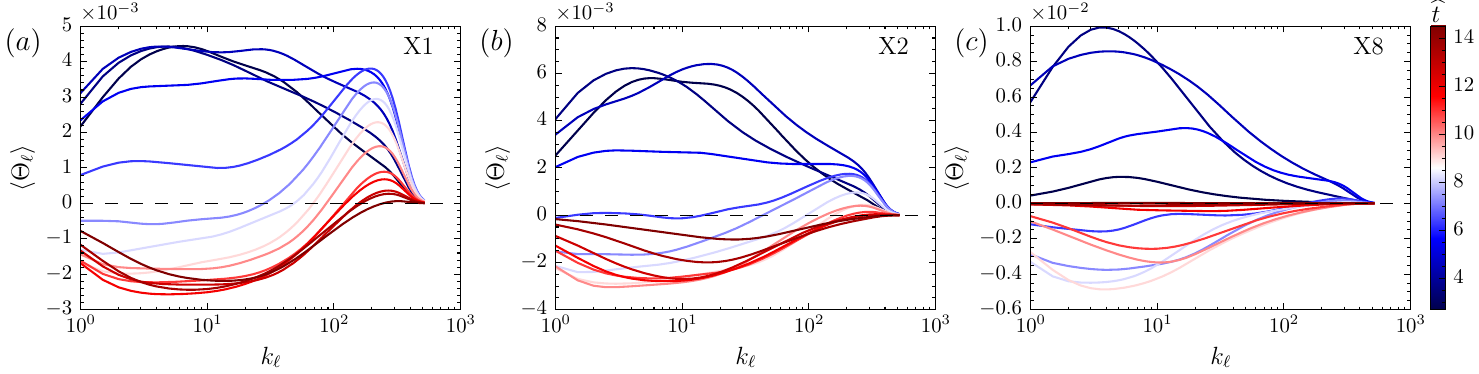} } 
\caption{Spatial average of the scalar variance transfer term $\Theta_\ell$ versus the filtering wavenumber $k_\ell=L_x/\ell$ at different time instants indicated by the colorbar. Panels (a),(b),(c) corresponds to simulation cases X1, X2, and X8, respectively. \label{fig:filtered_c_var}}
\end{minipage}
\end{figure}

Figure \ref{fig:filtered_c_var} presents the spatially averaged transfer term $\langle \Theta_\ell\rangle$ as a function of the filtering wavenumber $k_\ell=L_x/\ell$. The temporal evolution is illustrated by the color-code of the curves, from dark blue (early time) to dark red (late time). For Case X1 (panel a), the RT growth stage ($\widehat{t}\lesssim 5$, blue lines) exhibits positive $\langle \Theta_\ell\rangle$ across all wavenumbers. This indicates a forward cascade where scalar variance moves to smaller scales, driving the formation of fine-scale structures. During the turbulent emulsion stage ($5\lesssim \widehat{t}\lesssim 8$, light lines), the dynamics shift: $\langle \Theta_\ell\rangle$ becomes negative at large scales while remaining positive at small scales, representing simultaneous inverse transfer at large scales and forward transfer at small scales. By the phase segregation stage ($\widehat{t}\gtrsim 8$, red lines), the forward transfer region diminishes, and the flow is dominated by inverse transfer. For the higher–surface-tension cases X2 and X8 (panels b and c), the behavior is broadly similar, although the dual-cascade structure of $\langle \Theta_\ell\rangle$ in the turbulent emulsion stage becomes progressively less pronounced as $\sigma$ increases.

The scale-by-scale budget term in figure~\ref{fig:filtered_c_var} is presented in cumulative form as a function of the cutoff wavenumber $k_\ell=L_x/\ell$, corresponding directly to the filtered budget at each filter scale. To identify the dominant contributing scales more clearly, we also compute the logarithmic-scale density
$$\frac{d \langle \Theta_\ell\rangle}{d\log k_\ell}
= k_\ell \frac{d \langle \Theta_\ell\rangle}{d k_\ell}.$$
The resulting density plots are shown in figure~\ref{Appfig:cflux_density} of the Appendix. Positive values indicate an increase of $\langle\Theta_\ell\rangle$ contributed by scale $\ell$, providing a physically meaningful diagnostic of the scale-by-scale contribution to scalar-variance transfer. Analogous density representations for the kinetic-energy budgets are shown in Appendix figure~\ref{Appfig:flux_density}.

The scalar variance transfer term $\Theta_\ell$, as we demonstrate below, is intrinsically linked to the (filtered) tangential strain-rate, defined as $\overline{\mathcal{S}}_\ell=-\overline{\boldsymbol{n}}~\overline{\boldsymbol{n}}: \nabla\overline{\bu}$, where $\overline{\boldsymbol{n}} = \nabla \overline{c}_\ell/| \nabla \overline{c}_\ell|$ is the unit normal vector. Here, it is equal to the stretching term quantifying the rate of surface deformation, as demonstrated by the relation \citep{PoinsotVeynante,ZhaoLi25JFM}:
\begin{align}
    \overline{\mathcal{S}}_\ell = - \overline{\boldsymbol{n}}~\overline{\boldsymbol{n}} : \nabla\overline{\bu} = (\boldsymbol{\mathrm{I}} - \overline{\boldsymbol{n}}~\overline{\boldsymbol{n}}):\nabla\overline{\bu} =  \frac{1}{\widetilde{A}} \frac{D\widetilde{A}}{Dt}
\end{align}
where $\widetilde{A}$ represents local area density in the filtered scalar field, and $D/Dt$ represents the material derivative.The incompressibility condition is adopted in the derivation above. The connection between surface stretching $\overline{\mathcal{S}}_\ell$ and scalar variance transfer $\Theta_\ell$ is established via the nonlinear model (or gradient model). This approximation relies on the scale locality of the interactions and a Taylor series expansion of the filtered fields \citep{BorueOrszag98,Eyink06JFM,ZhaoLi25JFM}.

Assuming a Gaussian filtering kernel, the nonlinear model approximates the sub-scale flux $\overline{\tau}_\ell(c,\bu)$ as \citep{ZhaoLi25JFM}:
\begin{align}
    \overline{\tau}_\ell(c,\bu) \approx \frac{1}{12}\ell^2\nabla \overline{c}_\ell \cdot \nabla  \overline{\bu}
\end{align}
Consequently, $\Theta_\ell$ and $\overline{\mathcal{S}}_\ell$ are related by:
\begin{align} \label{eq:theta_K_relation}
\begin{split}
    \Theta_\ell &= -\nabla\overline{c}_\ell\cdot \overline{\tau}_\ell(c, \bu) \approx -\frac{1}{12}\ell^2 \nabla\overline{c}_\ell \nabla\overline{c}_\ell :\nabla\overline{\bu} \\
    &= -\frac{1}{12}\ell^2 \overline{\mathcal{S}}_\ell |\nabla\overline{c}_\ell|^2 \equiv \Theta^\mathrm{NL}_\ell
\end{split}
\end{align}

The link between scalar-variance transfer and interface stretching is physically rooted in the relationship between the filtering spectrum $\overline{E}_c$ and the interfacial area $A$. From the filtered scalar-variance budget,
\begin{equation}
\frac{1}{2}\frac{\partial}{\partial t}
\langle|\overline c_\ell|^2\rangle
=
-\langle\Theta_\ell\rangle ,
\end{equation}
together with the definition
$\overline E_c(k_\ell)=\frac{\partial}{\partial k_\ell}\langle|\overline c_\ell|^2\rangle$, 
we obtain
\begin{equation}
-\frac{\partial \langle\Theta_\ell\rangle}{\partial k_\ell}=
\frac{1}{2}
\frac{\partial \overline E_c(k_\ell)}{\partial t}.
\end{equation}
In the Porod regime, $\overline E_c \sim C A(t) k_\ell^{-2}$, this implies
\begin{equation}
\langle\Theta_\ell\rangle
\sim
\frac{C}{2}\frac{dA}{dt}k_\ell^{-1}
\sim
\ell \frac{dA}{dt}.
\end{equation}
Thus, scalar-variance transfer is governed by the rate of change of interfacial area, i.e. by interface stretching, $\frac{1}{A}\frac{dA}{dt}$. The nonlinear model above provides the corresponding local expression and gives the coefficient $1/12$ for the Gaussian filter.

\begin{figure}
\centering 
\begin{minipage}[b]{1.0\textwidth}  
\centering
\subfigure{\includegraphics[height=1.2in]{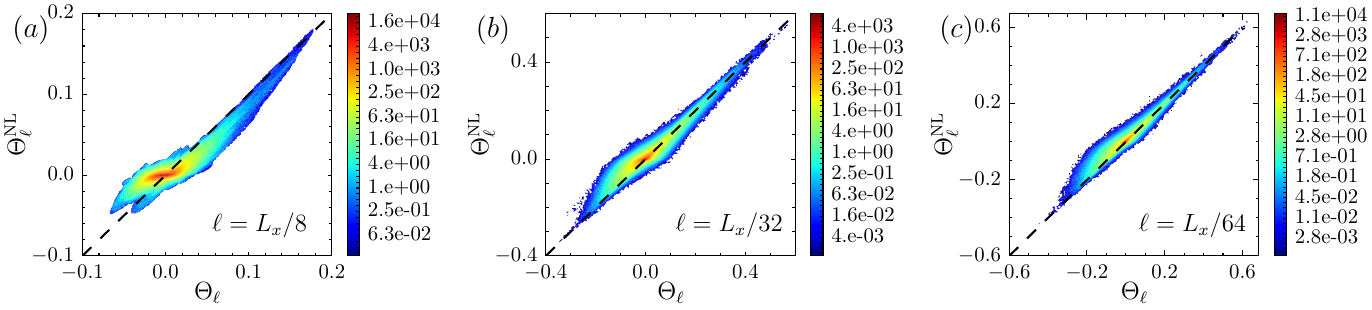} } 
\caption{Joint PDFs of the scalar variance transfer term $\Theta_\ell$ and the nonlinear model approximation $\Theta^\mathrm{NL}_\ell=-\frac{1}{12}\ell^2 |\nabla \overline{c}_\ell|^2 \overline{\mathcal{S}}_\ell$. The data correspond to simulation case X1 at time $\widehat{t}=6.3$. The panels represent filtering wavenumbers of (a) $k_\ell=8$, (b) $k_\ell=32$, and (c) $k_\ell=64$. The correlation coefficients for these scales are 0.81, 0.94, and 0.97, respectively.
\label{fig:stretch_theta_jpdf}}
\end{minipage}
\end{figure}

Figure \ref{fig:stretch_theta_jpdf} displays the joint PDFs of $\Theta_\ell$ and $\Theta^\mathrm{NL}_\ell$ for Case X1 at $\widehat{t}=6.3$, evaluated at three filter widths. The two terms show strong agreement with high correlation coefficients, particularly at small $\ell$. Similar correlations are found at other times and across different simulation cases. These results indicate that forward variance transfer ($\langle\Theta_\ell\rangle>0$) is physically associated with interface stretching at scales $\gtrapprox\ell$, whereas inverse transfer ($\langle\Theta_\ell\rangle<0$) corresponds to interface compression.

\subsection{Transfer of kinetic energy across scales}

\begin{figure}
\centering 
\begin{minipage}[b]{1.0\textwidth}  
\centering
\subfigure
{\includegraphics[height=2.2in]{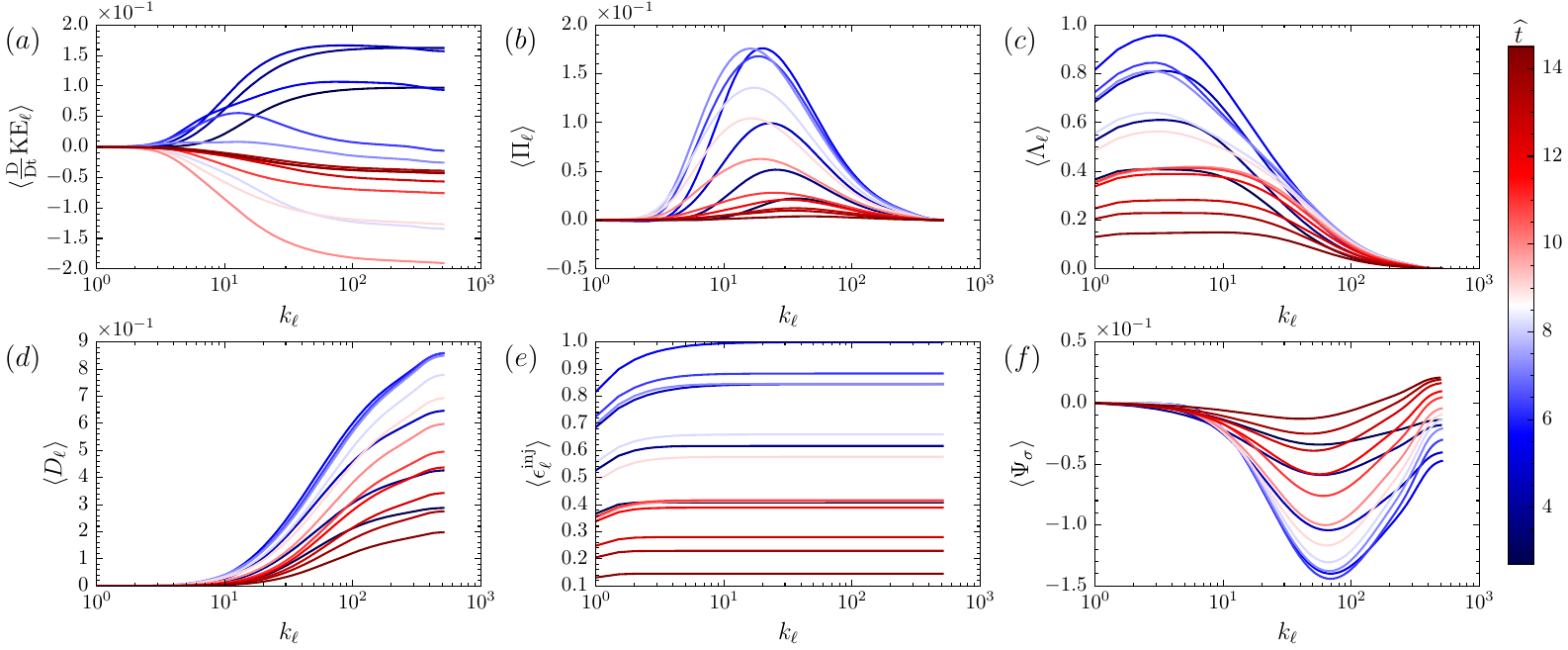} } 
\caption{Filtered kinetic energy budget terms from equation~(\ref{eq:filtered_KE_budgets}) for Case X1, plotted as functions of the filtering wavenumber $k_\ell$. The curves correspond to different time instants, color-coded according to the colorbar. All terms are normalized by the maximum unfiltered injection rate observed over the entire simulation duration. \label{fig:KE_cascade}}
\end{minipage}
\end{figure}

Figure \ref{fig:KE_cascade} illustrates the averaged filtered kinetic energy budgets (equation \ref{eq:filtered_KE_budgets}) as functions of the filtering wavenumber for Case X1; the other cases show qualitatively similar behavior. In panel (a), the time derivative of the filtered kinetic energy, $d\langle \mathrm{KE}_\ell \rangle/dt$, where  
$\mathrm{KE}_\ell \equiv \overline{\rho}_\ell |\widetilde{\bu}_\ell|^2/2$, is positive during the RT growth stage ($\widehat{t}\lesssim 5$, blue lines), indicating a net increase in kinetic energy. During the turbulent emulsion stage ($5\lesssim \widehat{t}\lesssim 8$, light-colored lines), the mean value becomes negative as dissipation exceeds the energy injection rate. Finally, in the phase segregation stage ($\widehat{t}\gtrsim 8$, red lines), it remains negative as kinetic energy continues to decay.

The deformation work $\Pi_\ell$ and baropycnal work $\Lambda_\ell$ (figure~\ref{fig:KE_cascade} b,c) act as cascade terms transferring kinetic energy across scales. As in 3-D miscible RT turbulence \citep{ZhaoAluie22JFM}, both are positive on average, indicating a net forward transfer. Their magnitudes increase during RT growth, peak in the turbulent-emulsion stage, and decay during segregation. The peak of $\langle\Pi_\ell\rangle$ shifts from $k_\ell \approx 40$ initially to $\approx 20$ near peak turbulence before returning toward $\approx 40$, whereas $\langle\Lambda_\ell\rangle$ remains concentrated at low wavenumbers ($k_\ell \approx 3$) and is typically larger. Both the injection and dissipation terms (figure~\ref{fig:KE_cascade} d,e) evolve in time similarly, with injection confined to the largest scales ($k_\ell<5$, negligible for $k_\ell\gtrsim10$) and dissipation dominated by high wavenumbers ($k_\ell \approx 100$).

Finally, panel (f) shows the filtered surface tension power. The unfiltered value ($k_\ell\to\infty$) is negative during the growth and emulsion stages, thus converting kinetic energy into surface energy via interface deformation, but becomes positive during phase segregation as the interface relaxes. The scale-by-scale contribution, indicated by the derivative $d\langle \Psi_{\sigma,\ell}\rangle/dk_\ell$, is negative at large-to-intermediate scales and positive at small scales throughout the simulation. The crossover wavenumber (where the derivative is zero) varies with time. Notably, this crossover wavenumber has been used in recent studies \citep{Crialesi23CP,CannonRosti24JFM} to define the Hinze scale, offering an alternative to the traditional definition in equation (\ref{eq:scales}) for homogeneous isotropic two-phase turbulence.

\begin{figure}
\centering 
\begin{minipage}[b]{1.0\textwidth}  
\centering
\subfigure{\includegraphics[height=1.5in]{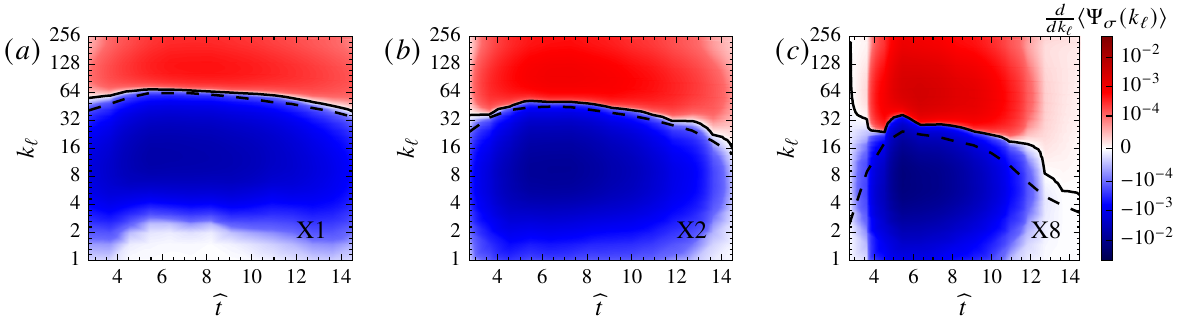} } 
\caption{Time-wavenumber evolution of the scale-by-scale surface tension power contribution, $d\langle \Psi_{\sigma}(k_\ell)\rangle/dk_\ell$, for Cases X1, X2, and X8, starting at $\widehat{t}=2.7$ (established turbulence). The filtering wavenumber is shown on a logarithmic scale. Solid lines indicate zero-crossings; dashed lines mark the Hinze scale (Eq.~\ref{eq:scales}).  \label{fig:surf_work_diagram}}
\end{minipage}
\end{figure}

While the two definitions of the Hinze scale are known to align in steady homogeneous isotropic turbulence \citep{Crialesi23CP,CannonRosti24JFM}, here we compare them in the context of anisotropic, non-stationary immiscible RT turbulence. Figure~\ref{fig:surf_work_diagram} presents time-wavenumber diagrams of the scale-by-scale surface tension contribution to kinetic energy, $d\langle \Psi_\sigma(k_\ell)\rangle/dk_\ell$, for cases X1, X2, and X8. In Case X1 (panel a), the diagram shows a blue region at low-to-intermediate wavenumbers, indicating where kinetic energy is converted into surface energy. Conversely, the red region at high wavenumbers indicates where surface energy returns to kinetic energy. The solid line marks the zero-crossing boundary between these regions, while the dashed line represents the traditional Hinze scale (equation \ref{eq:scales}). The close proximity of these lines confirms that the traditional Hinze definition effectively identifies the transition scale where surface tension switches from acting as a sink to a source.

For the high surface tension cases X2 and X8 (figure~\ref{fig:surf_work_diagram} b and c), the Hinze scale shifts to larger values, and the result of case X2 resembles that of case X1. In contrast for case X8, it exhibit larger   deviations between the two definitions. This discrepancy likely arises because high surface tension suppresses the formation of small-scale structures (as seen in the visualization in figure~\ref{fig:field_viz} (f)). Consequently, scalar turbulence and its hierarchy of length scales are not fully established, reducing the accuracy of the traditional Hinze definition in predicting the energetic crossover scale.

\begin{figure}
\centering 
\begin{minipage}[b]{1.0\textwidth}  
\centering
\subfigure{\includegraphics[height=1.2in]{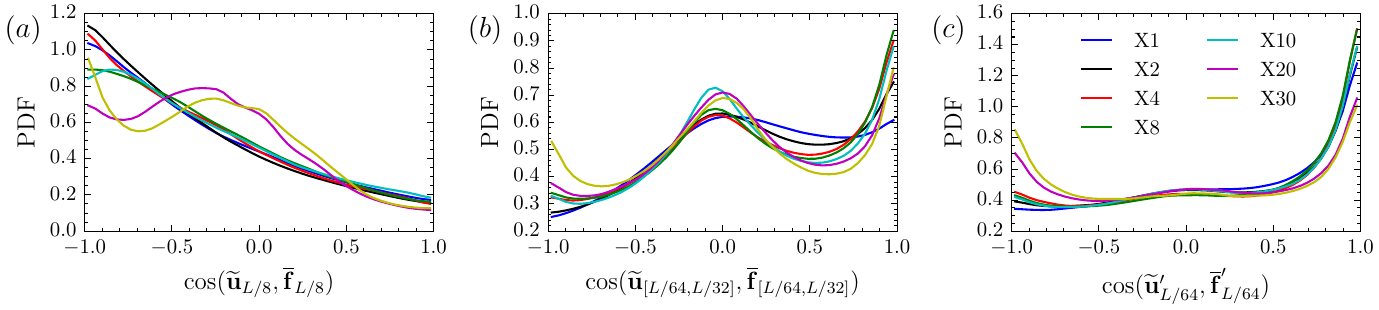} } 
\caption{PDFs of the alignment cosine between the velocity and surface tension fields at time $\widehat{t}=6$, evaluated at the interface. Panels show: (a) large scales ($\ell > L_x/8$); (b) intermediate scales ($L_x/64 < \ell < L_x/32$), where the band-pass filtered velocity $\widetilde{\bu}_{[\ell_1,\ell_2]} \equiv \widetilde{\bu}_{\ell_1} - \widetilde{\bu}_{\ell_2}$; and (c) small scales ($\ell < L_x/64$), where the residual velocity $\bu'_\ell \equiv \bu - \widetilde{\bu}_\ell$.  \label{fig:cos_uf}}
\end{minipage}
\end{figure}

The scale-dependent role of surface tension, which acts as either a sink or a source of filtered kinetic energy, is elucidated by the alignment cosine between the velocity and surface tension fields, shown in figure \ref{fig:cos_uf}. At large scales ($\ell>L_x/8$, panel a), the velocity and surface tension force tend to be anti-aligned, as evidenced by the probability density functions (PDFs) peaking at -1. At intermediate scales ($L_x/64 < \ell < L_x/32$, panel b), this anti-alignment diminishes, and the PDFs shift toward positive cosine values. Finally, at small scales ($\ell < L_x/64$, panel c), the PDFs peak at 1, indicating a preferential parallel alignment between the two fields.

These alignment trends result in negative surface tension power at large scales and positive value at small scales, as observed in figure \ref{fig:surf_work_diagram}. Physically, large scales in immiscible RT turbulence are dominated by inertial effects driven by energy injection and the cascade process. This causes large-scale interface distortion (area increase), meaning the velocity field works against surface tension, thereby converting kinetic energy into surface energy. In contrast, small scales are dominated by viscous effect and surface tension. Here, interface perturbations tend to relax toward equilibrium; this restoring force drives the small-scale velocity field, resulting in alignment between the force and velocity. Consequently, surface energy is converted back into kinetic energy. This phenomenon is most pronounced during the RT growth and turbulent emulsion stages, where energy injection and the cascade process are dominant.

\subsection{Connection between surface tension power and interface stretching} \label{sec:interface_stretching}

The budget equation (\ref{eq:filtered_csq_budgets}) for filtered scalar variance and its nonlinear model (\ref{eq:theta_K_relation}) identify $\Theta_\ell$ as the cascade term for scalar variance. Since this term governs the evolution of scalar variance, it is intrinsically linked to changes in interfacial area (stretching or compression) and, consequently, the total surface energy. Therefore, the scalar variance cascade $\Theta_\ell$ essentially reflects the energy change due to surface tension power.

This connection is quantified using the continuous surface force method \citep{Brackbill92JCP}, where the surface tension $\boldsymbol{f}^\sigma=\sigma\kappa\nabla c$. For unfiltered quantities, the surface tension power $\Psi_\sigma$ can be expanded as:
\begin{align}
    \begin{split}
        \Psi_\sigma &= \boldsymbol{f}^\sigma\cdot\bu = \sigma \kappa \bu\cdot\nabla c=\sigma \nabla\cdot\left(-\frac{\nabla c}{|\nabla c|}\right) \bu\cdot\nabla c\\        
        &= -\nabla\cdot\left(\sigma\frac{\nabla c}{|\nabla c|} \bu\cdot\nabla c - \sigma \bu |\nabla c|\right) + \frac{\sigma}{|\nabla c|} \nabla c \cdot \nabla \bu \cdot \nabla c \\
        &= \sigma \nabla\cdot\left[|\nabla c| \left(\boldsymbol{\mathrm{I}}-\mathrm{\boldsymbol{n}}\mathrm{\boldsymbol{n}}\right)\cdot\bu \vphantom{\frac{1}{1}}\right] - \sigma |\nabla c| \mathcal{S}
    \end{split} \label{eq:stretch_surf_relation}
\end{align}
where $\mathcal{S}=-\boldsymbol{n}\boldsymbol{n}:\nabla \bu$ equals the tangential strain-rate and $\boldsymbol{n} = \nabla c / |\nabla c|$ is the interface normal. In the final line, the first divergence term represents the transport of interfacial area density ($|\nabla c|$) by the tangential velocity field, while the second term is proportional to the interface stretching $\mathcal{S}$ in incompressible flows. Consequently, when integrated over an isolated bubble or droplet, the divergence term vanishes, leaving the surface tension power proportional to the interface stretching. Moreover, after volume averaging over the full domain, the right-hand side recovers the interfacial-area evolution, $\langle \mathbf{f}^{\sigma}\cdot \mathbf{u}\rangle = -\sigma\, dA/dt$, through the area-density $|\nabla c|$ budget, as shown in figure~\ref{fig:bulk_KE_budgets}(b).

The relation between surface-tension power and changes in interfacial area is well established \citep{DoddFerrante16JFM,VelaAvila2021JFM,CaladoBalaras24PRF}, both in integrated form and in local phase-field form. Here, we recast this relation in a local VOF-CSF formulation. Specifically, the surface-tension power is decomposed into a conservative transport term for the interfacial area density and a local surface-stretching contribution.

While derived for unfiltered fields, this relationship extends to coarse-grained fields with minor modifications. Specifically, the filtered surface tension force, $\overline{\boldsymbol{f}^\sigma}$, does not simply equal the product of filtered curvature and filtered scalar gradient (i.e., $\overline{\boldsymbol{f}}=\sigma \overline{\kappa \nabla c} \neq \sigma \overline{\kappa}~\nabla\overline{c}$); this discrepancy is typically handled via a subgrid scale model \citep{Labourasse07IJMF, Saeedipour19IJMF}. However, if we approximate the filtered work as $\Psi_{\sigma,\ell} \approx \sigma \overline{\kappa}_\ell \overline{\bu}_\ell\cdot \nabla\overline{c}_\ell$ (the calculated correlation coefficient between $\overline{\boldsymbol{f}^\sigma}$ and $\sigma \overline{\kappa}~\nabla\overline{c}$ exceeds 0.5 at all scales and is greater than 0.8 for scales smaller than $L/64$), the key conclusion remains that filtered surface tension power is closely related to the filtered interface stretching. More specifically, 
\begin{align} \label{eq:stretch_surf_filter}
    \Psi_{\sigma,\ell} \approx \sigma\overline{\kappa}_\ell \overline{\bu}_\ell\cdot \nabla\overline{c}_\ell = \sigma \nabla\cdot\left[|\nabla \overline{c}_\ell| \left(\boldsymbol{\mathrm{I}}-\mathrm{\boldsymbol{n}}\mathrm{\boldsymbol{n}}\right)\cdot\overline{\bu}_\ell \vphantom{\frac{1}{1}}\right] -  \sigma |\nabla \overline{c}|\overline{\mathcal{S}}_\ell
\end{align}
where the normal vector $\mathrm{\boldsymbol{n}}= \nabla\overline{c}_\ell/|\nabla\overline{c}_\ell|$ and $\overline{\mathcal{S}}_\ell \equiv -\overline{\boldsymbol{n}} \cdot \nabla\overline{\bu}_\ell\cdot \overline{\boldsymbol{n}}$ equals the coarse-grained tangential strain-rate, which equals to the interface stretch rate in incompressible flows.

Given the connection between surface tension power and interface stretching, the scale-dependent role of surface tension is also manifested in the sign of the mean stretching filtered at different length scales. By recasting the stretching term into the eigen-basis of the filtered strain-rate tensor, we obtain:
\begin{align}
\begin{split}
     \overline{\mathcal{S}}_\ell |\nabla\overline{c}_\ell|^2 &= -\nabla\overline{c}_\ell \cdot \nabla\overline{\bu}_\ell\cdot \nabla\overline{c}_\ell=-\nabla\overline{c}_\ell \cdot \overline{\boldsymbol{\mathrm{S}}}_\ell\cdot \nabla\overline{c}_\ell \\
     &=-\nabla\overline{c}_\ell\cdot (\lambda_\alpha \boldsymbol{\widehat{e}}_\alpha\boldsymbol{\widehat{e}}_\alpha+\lambda_\beta \boldsymbol{\widehat{e}}_\beta\boldsymbol{\widehat{e}}_\beta+\lambda_\gamma \boldsymbol{\widehat{e}}_\gamma\boldsymbol{\widehat{e}}_\gamma)\cdot\nabla\overline{c}_\ell\\
     &= -\lambda_\alpha \cos(\nabla\overline{c}_\ell,\boldsymbol{\widehat{e}}_\alpha)^2 -\lambda_\beta \cos(\nabla\overline{c}_\ell,\boldsymbol{\widehat{e}}_\beta)^2 -\lambda_\gamma \cos(\nabla\overline{c}_\ell,\boldsymbol{\widehat{e}}_\gamma)^2  
\end{split} \label{eq:stretch_eigen_decomp}
\end{align}
where $\lambda_\alpha > \lambda_\beta > \lambda_\gamma$ are the eigenvalues of the filtered strain-rate tensor $\overline{\boldsymbol{\mathrm{S}}}_\ell \equiv \frac{1}{2}(\nabla \overline{\bu}+\nabla\overline{\bu}^T)$, and $\boldsymbol{\widehat{e}}_\alpha, \boldsymbol{\widehat{e}}_\beta, \boldsymbol{\widehat{e}}_\gamma$ are the associated eigenvectors. In incompressible flows, $\lambda_\alpha > 0$ (extensive) and $\lambda_\gamma < 0$ (compressive). Consequently, the sign of the mean stretching depends on the alignment of the scalar gradient with these eigenvectors. Alignment with the extensive eigenvector ($\nabla\overline{c}_\ell \parallel \widehat{\boldsymbol{e}}_\alpha$) yields a negative contribution, while alignment with the compressive eigenvector ($\nabla\overline{c}_\ell \parallel \widehat{\boldsymbol{e}}_\gamma$) yields a positive contribution.

\begin{figure}
\centering 
\begin{minipage}[b]{1.0\textwidth}  
\centering
\subfigure{\includegraphics[height=2.4in]{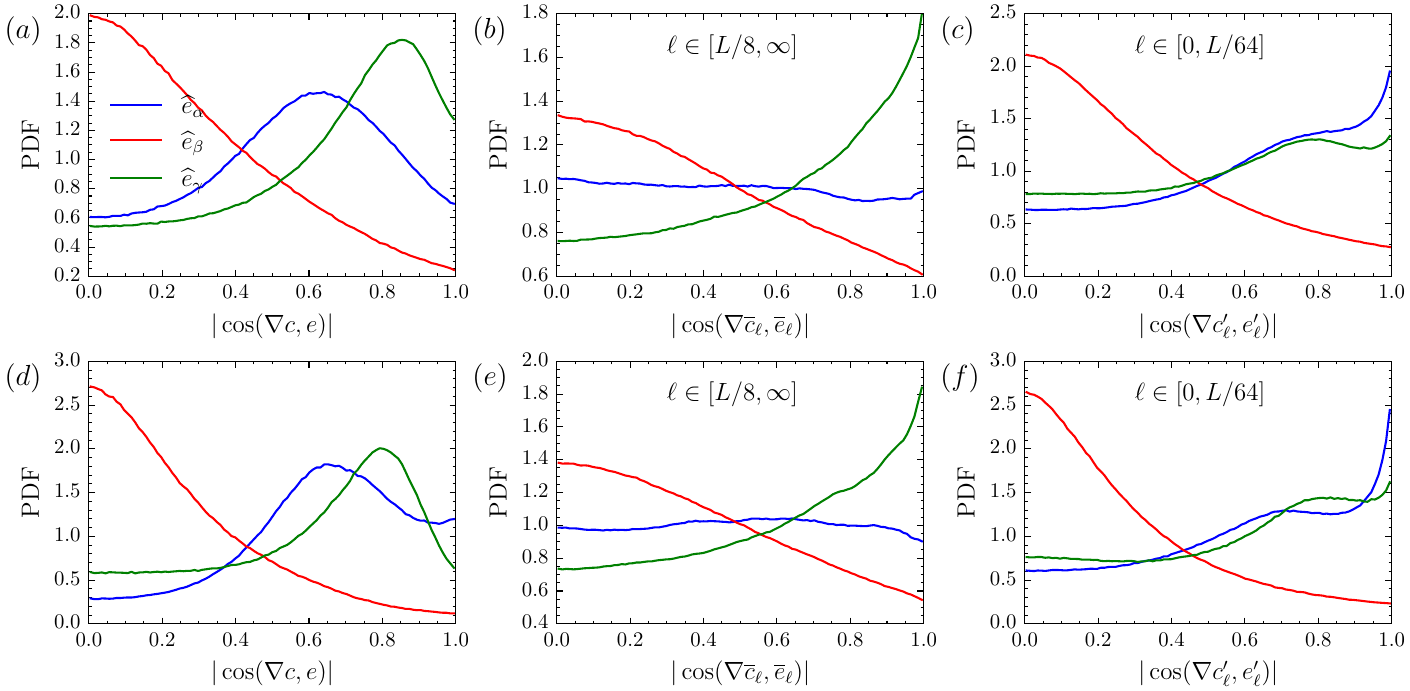} } 
\caption{Alignment between the scalar gradient and the eigenvectors of the strain-rate tensor for simulation case X1 at non-dimensional times $\widehat{t}=6$ (a-c) and $\widehat{t}=13.6$ (d-f). The columns correspond to: (a, d) unfiltered fields; (b, e) large-scale fields (scales $> L_x/8$); and (c, f) small-scale fields (scales $< L_x/64$). The high-pass filter is defined as $c'_\ell\equiv c-\overline{c}_\ell$ and $e_l'$ is the eigenvector of the high-pass strain-rate tensor.
 \label{fig:align_dc_du}}
\end{minipage}
\end{figure}

In miscible RT turbulence \citep{ZhaoLi25JFM}, the scalar gradient preferentially aligns with the compressive eigenvector ($\widehat{\boldsymbol{e}}_\gamma$), favoring positive interface stretching. For the immiscible RT turbulence studied here, the alignment is shown in Figure~\ref{fig:align_dc_du}. For the unfiltered fields (panels a,d), $\nabla c$ is generally perpendicular to the intermediate eigenvector $\widehat{\boldsymbol e}_\beta$ and preferentially aligned with the compressive eigenvector $\widehat{\boldsymbol e}_\gamma$; the mean angles between $\nabla c$ and $\widehat{\boldsymbol e}_\gamma$ and $\widehat{\boldsymbol e}_\alpha$ are approximately $35^\circ$ and $55^\circ$, respectively.

At large scales ($\ell>L_x/8$, panels b, e), the preferential alignment of $\nabla\overline{c}_\ell$ with $\widehat{\boldsymbol{e}}_\gamma$ is strong, indicating that mean interface stretching is positive. In contrast, for high-pass filtered fields representing small scales ($\ell<L_x/64$, using residual fields), the trend reverses that the scalar gradient preferentially aligns with the extensive eigenvector $\widehat{\boldsymbol{e}}_\alpha$, leading to negative interface stretching. These observations are consistent with the surface tension power depicted in figures~\ref{fig:surf_work_diagram} and \ref{fig:cos_uf}. At late time (panels d-f), the alignment statistics remains similar. Since surface tension power is negatively proportional to stretching (equation \ref{eq:stretch_surf_relation}), positive stretching at large scales corresponds to the sink role of surface tension in kinetic energy budgets, while negative stretching at small scales corresponds to the source role. This confirms the link between scalar variance transfer, interface stretching, and surface tension power. We will further assess the validity of equation~(\ref{eq:stretch_surf_relation}) when analyzing bubble and droplet dynamics in the following section.

\section{Bubble and droplet statistics \label{sec:bub_drop_stat}}

\begin{figure}
\centering 
\begin{minipage}[b]{1.0\textwidth}  
\centering
\subfigure{\includegraphics[height=4in]{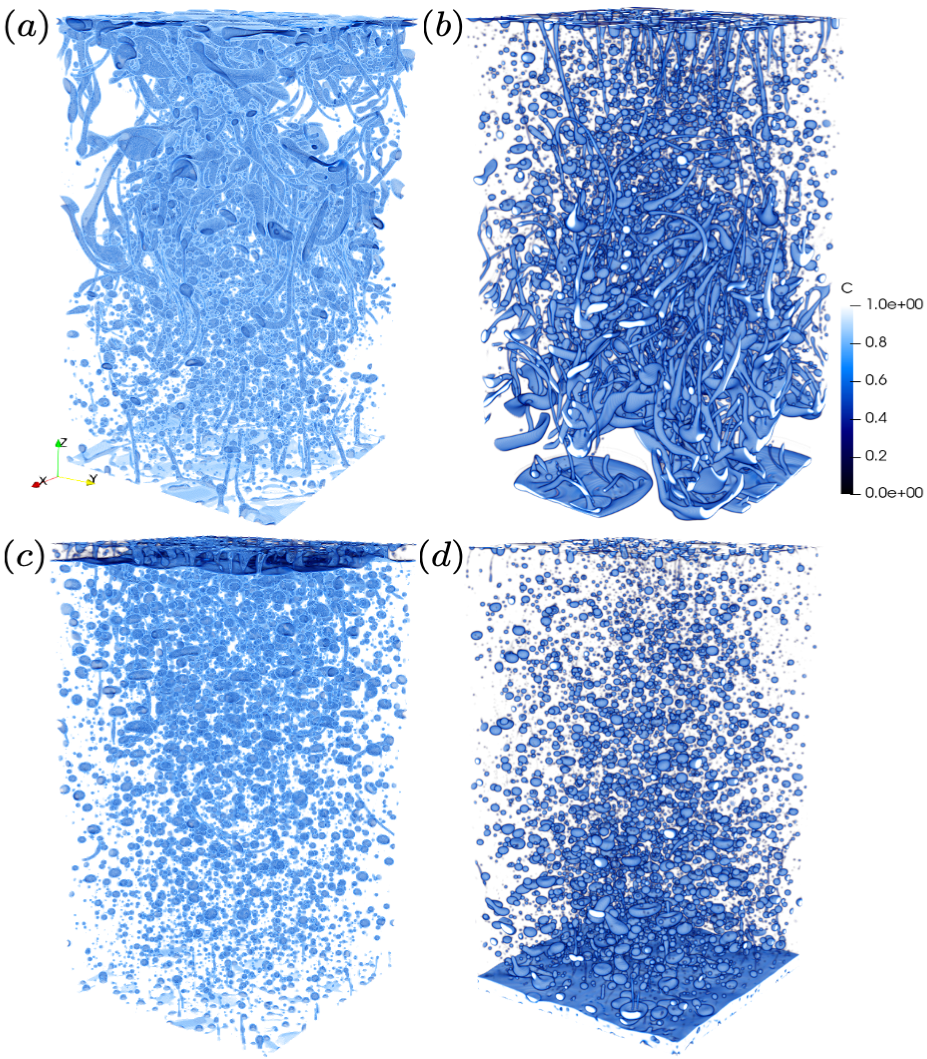} } 
\caption{Color function fields for cases X1$\Upphi 95$ (a, c) and X1$\Upphi 05$ (b, d) at non-dimensional times $\widehat{t}=6.5$ (a, b) and $\widehat{t}=12$ (c, d).
\label{fig:bubble_drop_viz}}
\end{minipage}
\end{figure}

During the turbulent emulsion and segregation stages, the immiscible RT turbulent flow contains isolated bubbles and droplets over a broad range of length scales. They significantly increase the total interfacial area, enhance the mixing of the resulting emulsion, and modulate both interface stretching and the scalar and kinetic energy budgets. To investigate the kinematics and dynamics of these structures, we performed simulations with heavy fluid volume fractions of $\Upphi=0.95$ (Case X1$\Upphi 95$) and $\Upphi=0.05$ (Case X1$\Upphi 05$). These conditions produce rich rising bubbles and sinking droplets, respectively, as the dispersed phase, allowing us to isolate and study their behaviors individually.
Figure~\ref{fig:bubble_drop_viz} visualizes the scalar field for these cases at non-dimensional times $\widehat{t}=6.5$ and $\widehat{t}=12$. At $\widehat{t}=6.5$, both the bubbles (panel a) and droplets (panel b) exhibit a mix of small spherical shapes and long filaments elongated along the vertical direction. By $\widehat{t}=12$, however, the morphology shifts that the number of long filamentary structures is significantly reduced, and the remaining bubbles and droplets are predominantly spherical. For statistical purpose, we hereafter denote the cases within the time range $5.5<\widehat{t}<6.5$ as `early time' and those within $11<\widehat{t}<12$ as `late time'. Numerically, isolated bubbles and droplets are identified using a 3-D two-pass connected-component algorithm, augmented with Union-Find and a decision tree \citep{RosenfeldPflatz68,cc3d_soft}.

\subsection{Bubbles and droplets distribution and morphology}
\begin{figure}
\centering 
\begin{minipage}[b]{1.0\textwidth}  
\centering
\subfigure{\includegraphics[height=1.8in]{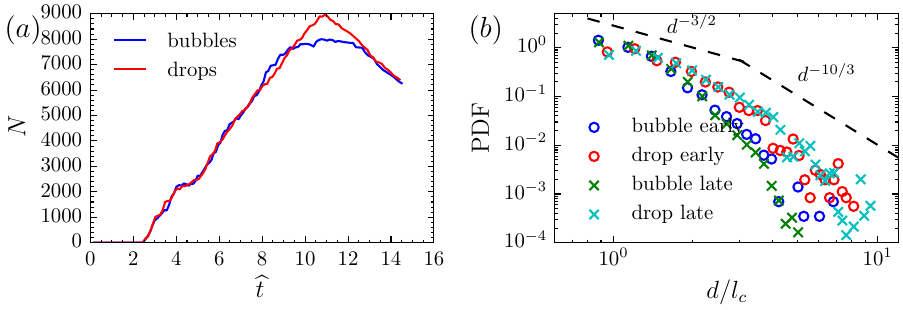} } 
\caption{(a) Time evolution of the total number of isolated bubbles and droplets for cases X1$\Upphi 95$ (labeled as bubbles) and X1$\Upphi 05$ (labeled as drops). (b) Droplet and bubble size distributions at $5.5<\widehat{t}<6.5$ (early time) and $11<\widehat{t}<12$ (late time).  Two dashed lines indicate reference scalings of $d^{-10/3}$ and $d^{-3/2}$, and $l_c\equiv \sqrt{\sigma/(\rho_h-\rho_l)/g}$ is the capillary scale.
 \label{fig:ndrops_dsd}}
\end{minipage}
\end{figure}

Figure~\ref{fig:ndrops_dsd} depicts the evolution of isolated bubbles and droplets for Cases X1$\Upphi 95$ and X1$\Upphi 05$. Panel (a) shows the total number of isolated structures. We observe that the dispersed phase begins to break up due to instability after $\widehat{t}>2$. Subsequently, the number of structures increases nearly linearly, reaches a peak at $\widehat{t} \approx 11$, and then decreases linearly. This initial linear growth is attributed to shear stripping driven by the Kelvin-Helmholtz instability, which continuously sheds small elements from the boundaries of large-scale primary structures (RT spikes and bubbles).

Figure~\ref{fig:ndrops_dsd}(b) shows the size distributions of bubbles (case X1$\Upphi95$) and droplets (case X1$\Upphi05$) at early ($5.5<\widehat{t}<6.5$) and late ($11<\widehat{t}<12$) times. Droplets tend to have larger equivalent diameters, $d \equiv (6V/\pi)^{1/3}$ (with $V$ the connected-element volume), consistent with the heavier phase being less susceptible to breakup than the lighter bubbles. The time evolution also differs between the two cases. For bubbles, the small-diameter portion of the distribution changes little between early and late times, whereas the large-diameter tail is reduced at late times, indicating a larger population of big bubbles during the initial RT growth that later fragments. By contrast, the droplet size distribution varies only weakly from early to late times, with changes mainly confined to the largest structures.

For reference, two power-law scalings are plotted in figure~\ref{fig:ndrops_dsd}(b): $d^{-3/2}$ and $d^{-10/3}$, representing coalescence-dominated and breakup-dominated regimes, respectively \citep{Crialesi22JFM,CannonRosti24JFM}. The droplet distribution follows these scalings over a limited inertial range. This consistency with classical Kolmogorov–Hinze theory suggests that droplet fragmentation is primarily driven by dynamic pressure fluctuations in the inertial subrange. Conversely, the bubble distribution decays more steeply and deviates from these standard power laws. This behavior is attributed to the light bubbles experiencing stronger buoyancy-induced shear and wake instabilities. These forces accelerate the disintegration of large structures via rapid surface stripping, resulting in the observed steeper decay in the size distribution.

\begin{figure}
\centering 
\begin{minipage}[b]{1.0\textwidth}  
\centering
\subfigure{\includegraphics[height=1.8in]{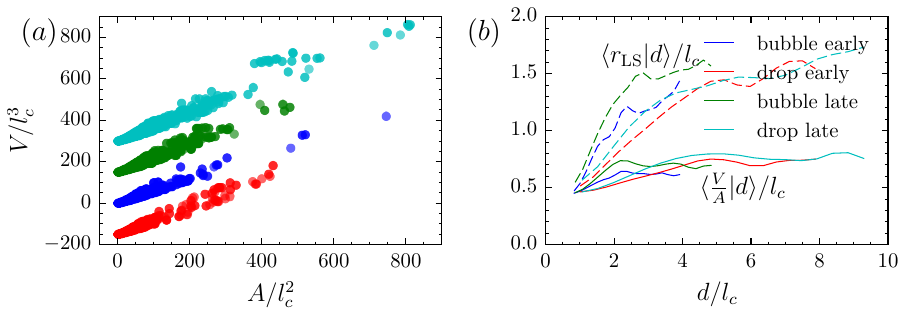} } 
\caption{(a) Scatter plot of normalized area versus normalized volume for isolated bubbles and droplets at early ($5.5<\widehat{t}<6.5$) and late ($11<\widehat{t}<12$) times. For clarity, the red, green, and cyan points are vertically shifted by -150, 150, and 300 units, respectively. (b) The mean volume-to-surface ratio, $\langle V/A|d\rangle$, and the mean maximum absolute signed distance, $\langle r_\mathrm{LS}|d\rangle$, conditioned on the equivalent diameter $d$. All quantities are normalized by the capillary scale, $l_c$.
\label{fig:vol_area_diam}}
\end{minipage}
\end{figure}

Beyond size distribution, we analyze the shape and morphology of the isolated bubbles and droplets. Figure~\ref{fig:vol_area_diam}(a) displays a scatter plot of the normalized area versus volume. Regardless of the evolution time (early vs. late), the data collapses onto a linear trend with similar slopes for bubbles and droplets. This implies that the volume-to-area ratio $V/A$ remains constant for bubbles and droplets across a wide range of sizes.
This ratio is further quantified in Figure~\ref{fig:vol_area_diam}(b), which plots the conditional mean $\langle V/A|d\rangle$ against the equivalent diameter $d$. For structures larger than $d/l_c \gtrsim 3$, the ratio plateaus, reaching an asymptotic value of approximately $0.6-0.8~l_c$. A constant $V/A$ ratio suggests that these large structures adopt a filamentary geometry where the length increases while the cross-section remains relatively constant. Approximating these filaments as cylinders, the cross-sectional diameter ($d_\mathrm{cross}$) relates to the volume-to-area ratio via $d_\mathrm{cross} \approx 4(V/A)$. Based on the observed asymptote, this predicts a filament thickness of $d_\mathrm{cross} \approx 2.4-3.2~ l_c$.

To validate this morphology, we analyze the maximum absolute signed distance within each element, $r_\mathrm{LS}\equiv \max(|d_\mathrm{sign}|)$, where $d_\mathrm{sign}$ is the signed distance to the interface and the maximum is taken over the discrete bubbles or droplets. Thus the quantity $r_\mathrm{LS}$ serves as a proxy for the cross-sectional radius. Figure~\ref{fig:vol_area_diam}(b) shows that the conditional mean $\langle r_\mathrm{LS}|d\rangle/l_c$ asymptotes to a value of $1.4-1.5$. This corresponds to a cross-sectional diameter of $2.8-3.0~l_c$, which is in good agreement with the prediction derived from the $V/A$ ratio. These results confirm that large structures in immiscible RT turbulence are predominantly filamentary, maintaining a characteristic thickness of approximately three capillary lengths.

\begin{figure}
\centering 
\begin{minipage}[b]{1.0\textwidth}  
\centering
\subfigure{\includegraphics[height=3in]{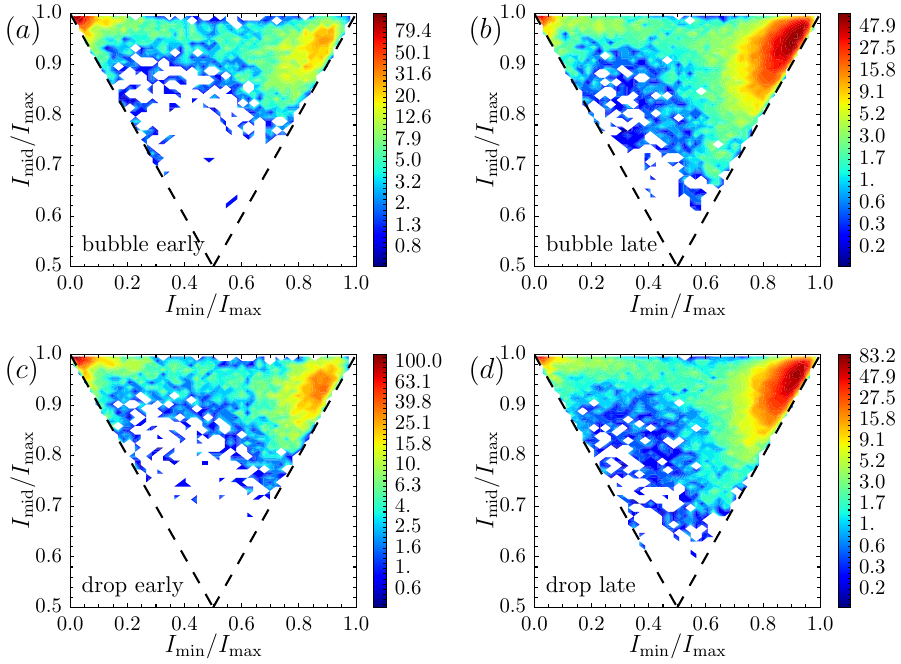} } 
\caption{Joint PDFs of the middle-to-maximum and minimum-to-maximum moment of inertia ratios. Panels (a, b) show the bubble case (X1$\Upphi 95$), while panels (c, d) show the droplet case (X1$\Upphi 05$). The left column (a, c) corresponds to early time ($5.5<\widehat{t}<6.5$), and the right column (b, d) corresponds to late time ($11<\widehat{t}<12$).
\label{fig:moment_inertia_jPDF}}
\end{minipage}
\end{figure}

To corroborate the filamentary nature of the large discrete elements, we analyze their three principal moments of inertia ($I_\mathrm{min} \le I_\mathrm{mid} \le I_\mathrm{max}$). Figure~\ref{fig:moment_inertia_jPDF} displays the joint PDFs of the ratios $I_\mathrm{min}/I_\mathrm{max}$ versus $I_\mathrm{mid}/I_\mathrm{max}$. These distributions are mathematically confined within a triangular region (note that $I_\mathrm{min}+I_\mathrm{mid}\geq I_\mathrm{max}$) where the vertices represent distinct morphologies: the top-left corresponds to filaments ($I_\mathrm{min} \ll I_\mathrm{mid} \approx I_\mathrm{max}$), the top-right to spheres ($I_\mathrm{min} \approx I_\mathrm{mid} \approx I_\mathrm{max}$), and the bottom to pancake-like structures ($I_\mathrm{min} \approx I_\mathrm{mid} \approx 0.5 I_\mathrm{max}$).

At early times ($5.5<\widehat{t}<6.5$), both the bubble (panel a) and droplet (panel c) cases exhibit a high concentration near the top-left corner. This indicates a dominance of filamentary structures, confirming the geometric results in figure~\ref{fig:vol_area_diam}. While spherical structures (top-right) are present, they remain secondary to the filaments during this phase.
In contrast, at late times ($11<\widehat{t}<12$, panels b and d), the region of highest probability shifts toward the top-right corner. This transition marks the emergence of numerous small spherical elements, consistent with the breakup mechanisms visualized in figure~\ref{fig:bubble_drop_viz}. Concurrently, the intensity at the top-left corner decreases slightly, reflecting a reduction in the relative population of filaments. Pancake structures appear primarily at late times, but their numbers remain statistically insignificant compared to filaments and spheres.

\begin{figure}
\centering 
\begin{minipage}[b]{1.0\textwidth}  
\centering
\subfigure{\includegraphics[height=3.4in]{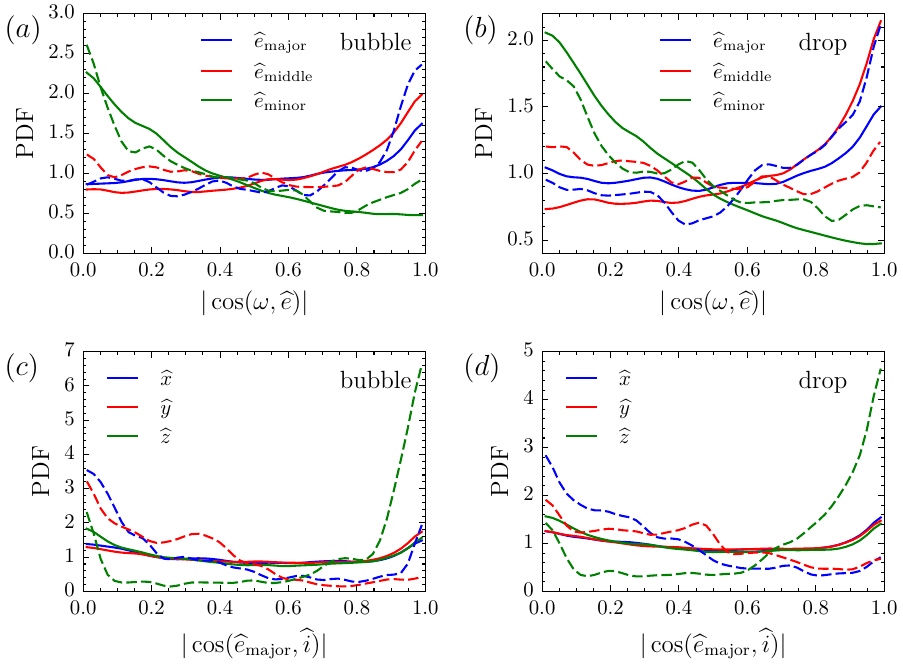} } 
\caption{Alignment of principal axes for isolated bubbles (case X1$\Upphi 95$; panels a, c) and droplets (case X1$\Upphi 05$; panels b, d) at late time ($11<\widehat{t}<12$). Panels (a, b) show the alignment with mean vorticity, while panels (c, d) show the alignment of the major axis with the three coordinate axes. Solid lines show all objects, while dashed lines correspond to long filamentary objects (moment of inertia ratio $I_\mathrm{min}/I_\mathrm{max}<0.15$).
\label{fig:drop_inertia_align}}
\end{minipage}
\end{figure}

We characterize the orientation of isolated bubbles and droplets using the principal axes of their moment-of-inertia tensor. Geometrically, the major, middle, and minor axes correspond to the minimum ($I_\mathrm{min}$), intermediate ($I_\mathrm{mid}$), and maximum ($I_\mathrm{max}$) moments of inertia, respectively. Figure~\ref{fig:drop_inertia_align} quantifies the orientation of these structures by analyzing the alignment of their principal axes with the mean vorticity vector (over each discrete element) and the fixed coordinate axes.
Panel (a) illustrates the alignment with the mean vorticity for the bubble case. The solid lines represent the full population, while the dashed lines represent the subset of filamentary structures ($I_\mathrm{min}/I_\mathrm{max}<0.15$). In the full statistics, the vorticity vector tends to be perpendicular to the minor axis, with a broad probability of aligning with both the major and middle axes. However, focusing on the filamentary subset reveals a strong preferential alignment between the major axis and the local vorticity. This behavior is primarily driven by vortex stretching: as vortex tubes stretch along their rotation axes, entrained elements are passively elongated in the same direction. This alignment also enhances stability, as filaments aligned with the stretching direction are less susceptible to immediate breakup by shear instabilities.

Panel (c) of figure~\ref{fig:drop_inertia_align} examines the alignment of the major axis with the coordinate system. The full population displays no preferential alignment, as the statistics are dominated by numerous small, spherical structures that lack a distinct orientation bias. In contrast, the filamentary subset (dashed lines) aligns preferentially with the vertical axis. This confirms that the elongated structures are generated by the vertical growth inherent to the RT instability.
Finally, panels (b) and (d) show that the droplet case mirrors the bubble behavior. Long filamentary droplets align simultaneously with the local mean vorticity and the global vertical direction.

\subsection{Bubbles and droplets dynamics}

\begin{figure}
\centering 
\begin{minipage}[b]{1.0\textwidth}  
\centering
\subfigure{\includegraphics[height=2.6in]{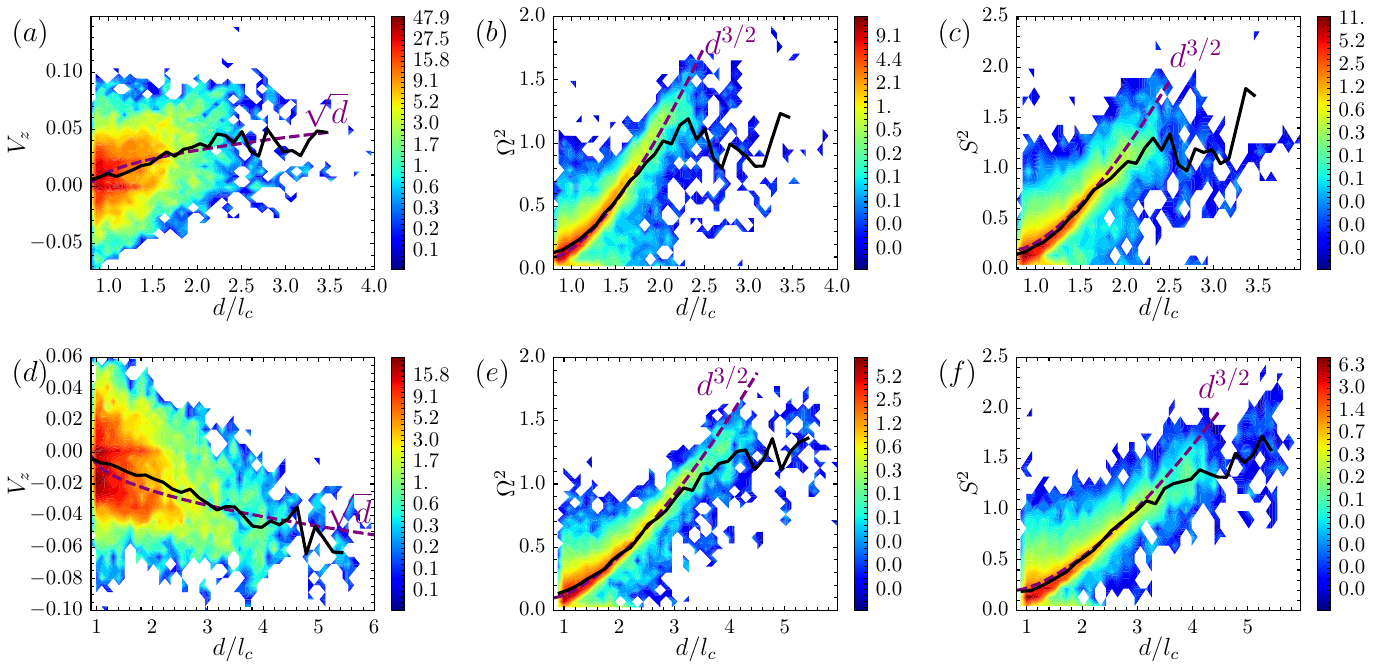} } 
\caption{Joint PDFs of the mean vertical velocity (a, d), mean enstrophy (b, e), and mean squared strain-rate (c, f) averaged over isolated structures, plotted against the equivalent diameter normalized by capillary length. Panels (a–c) correspond to the bubble case (X1$\Upphi 95$) and (d–f) to the droplet case (X1$\Upphi 05$) at late time ($11<\widehat{t}<12$). Solid black lines denote the conditional means. Purple dashed lines indicate theoretical scalings: the drag-buoyancy balance $\sim d^{1/2}$ in (a, d), and the dissipation scaling $\sim d^{3/2}$ in (b, c, e, f). 
\label{fig:bub_drop_ens_dissip}}
\end{minipage}
\end{figure}

The dynamics of discrete bubbles and droplets are intrinsically linked to their geometry and morphology. Figure~\ref{fig:bub_drop_ens_dissip} presents the joint PDFs of the element-averaged vertical velocity, enstrophy, and squared strain-rate plotted against the equivalent diameter.

We first examine the vertical velocity in panels (a,d). As expected for RT flows, the conditional mean vertical velocity is positive for bubbles (panel a) and negative for droplets (panel d), corresponding to rising and sinking motions, respectively. To explain the observed mean magnitude in solid lines, we employ a drag-buoyancy balance model. Assuming an isolated spherical element moving at terminal velocity, the drag force balances the buoyancy force:
\begin{align}
    \frac{1}{2} C_d \rho_{amb} V_z^2 A = (\rho_h-\rho_l)Vg,
\end{align}
where $C_d \approx 0.5$ is the drag coefficient for a sphere, $\rho_{amb}$ is the density of the ambient fluid, and $V$ and $A$ are the element's volume and cross-sectional area. This relation implies that the terminal velocity scales as $V_z \propto \sqrt{d}$.

However, this simple scaling breaks down for very small elements below the capillary scale, where the buoyancy force is negligible, and motion is dominated by ambient turbulence. To account for this, we introduce a modified scaling that incorporates a minimum size cutoff $l_\mathrm{cutoff}$ ($\sim l_c$, the capillary length) and a coefficient $c_0$ to represent turbulence modulation:
\begin{align}
    V_z = c_0 \sqrt{\frac{4}{3C_d} \frac{\rho_h-\rho_l}{\rho_{amb}} g (d-l_\mathrm{cutoff})}.
\end{align}
The dashed lines in Figure~\ref{fig:bub_drop_ens_dissip}(a, d) compare this model with the simulation data. Using a coefficient of $c_0=0.35$, the model shows good agreement with the conditional mean vertical velocity for both bubbles and droplets.

The analysis of element-averaged mean enstrophy and squared strain-rate (panels b, c, e, f) draws upon the inertial dynamics established in panels (a,d). The vertical velocity scaling, $\langle u_z \rangle \sim d^{1/2}$, indicates an inertial regime where turbulent drag balances buoyancy. Under Kolmogorov scaling, the energy dissipation rate is given by $\sim U^3/L$. Since the characteristic velocity grows as $U \sim d^{1/2}$ while the length scale $L$ remains effectively constant (limited to $\approx 3l_c$ by surface tension, see Figure \ref{fig:vol_area_diam}), the dissipation rate is predicted to scale as $d^{3/2}$. Thus the mean enstrophy and squared strain-rate over discrete structures follows this scaling. This prediction (purple dashed lines in figure~\ref{fig:bub_drop_ens_dissip} b,c,e,f) agrees well with the observed conditional mean profiles (black lines). Physically, this confirms that larger bubbles and droplets sustain more intense internal shear due to the combination of higher terminal velocities and fixed, capillary-limited widths.

\begin{figure}
\centering 
\begin{minipage}[b]{1.0\textwidth}  
\centering
\subfigure{\includegraphics[height=3in]{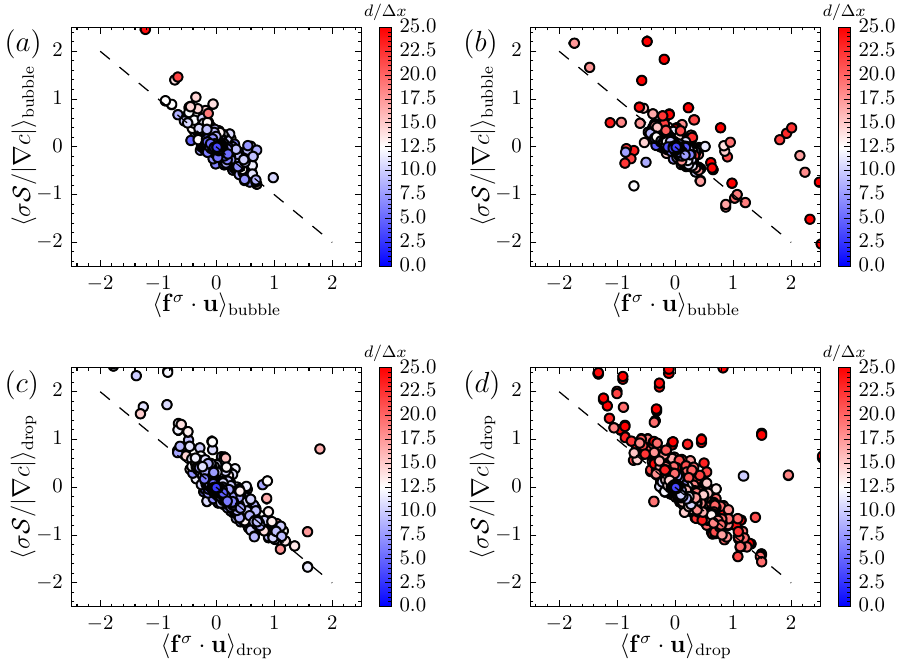} } 
\caption{Scatter plots of surface tension power versus surface stretching, integrated over individual bubbles (a, b) and droplets (c, d). Panels (a, c) represent early time ($5.5<\widehat{t}<6.5$), while panels (b, d) represent late time ($11<\widehat{t}<12$).
\label{fig:surfEne_scatter}}
\end{minipage}
\end{figure}

Finally, we verify the theoretical relationship between surface tension power and interface stretching derived in equation~(\ref{eq:stretch_surf_relation}). Figure~\ref{fig:surfEne_scatter} presents scatter plots of the element-averaged surface tension power, $\langle \boldsymbol{f}^\sigma\cdot\bu\rangle$, against the normalized interface stretching, $\langle \sigma\mathcal{S}~|\nabla c|\rangle$. Element averaging is essential here, as it eliminates the spatial transport contribution, thereby isolating the direct coupling between surface tension power and stretching.
Across all regimes, including bubbles and droplets, and at both early and late times, the data collapses well onto the diagonal line $y=-x$, providing strong numerical confirmation of the proposed balance. Some deviations are observed at late times (panels b and d), likely attributable to the dense packing of elements. In such configurations, the element-averaged divergence term may become non-negligible, resulting in slight statistical discrepancies. Nonetheless, the robust correlation $\langle \boldsymbol{f}^\sigma\cdot\bu\rangle \approx -\langle \sigma\mathcal{S} ~|\nabla c|\rangle$ confirms that surface tension power is fundamentally linked to interface stretching. This validates the physical framework proposed in Section~\ref{sec:scale_decomp}, linking surface tension power directly to the transfer of scalar variance across scales.

While demonstrated here for RT turbulence, the above mechanism describes a fundamental kinematic relationship that should be generally applicable to immiscible turbulent flows. 
This insight holds promise for the development of large-eddy-simulation (LES) models. In LES, where subgrid interfacial scales are often unresolved, this relation suggests that the transfer of kinetic energy to surface energy can be parameterized as a function of the resolved strain rate. By linking hydrodynamic straining directly to the generation of interfacial area, physics-based subgrid closures can be constructed. Such models would accurately capture the ``energy sink" effect of droplet formation and interface wrinkling without requiring explicit resolution of the complex subgrid geometry.

\section{Conclusions \label{sec:conclude}}

In this work, we investigated the interfacial dynamics and energy transfer mechanisms in immiscible two-phase Rayleigh-Taylor turbulence. By varying the surface tension coefficient $\sigma$, we analyzed the impact of capillarity on the energy budget, the cascade process, and the behaviour of discrete dispersed elements.

For the global dynamics, we identified three distinct stages in the RT evolution: the linear growth stage, the turbulent emulsion stage, and the phase segregation stage. The temporal evolution of the interfacial area can be estimated from the ratio of the mixed volume to the Hinze scale, with the Hinze scale being bounded from below by the capillary scale. The global evolution of energy budgets exhibits self-similarity with respect to surface tension: the flow duration scales as $\sigma^{-1/4}$, the maximum kinetic energy as $\sigma^{1/2}$, and the maximum dissipation rate as $\sigma^{1/4}$. Consequently, the temporal evolution of kinetic, potential, internal, and surface energies collapse onto same curves upon rescaling. At small scales, the role of surface tension transitions depending on the capillary strength. At low $\sigma$, it acts primarily as an enstrophy sink, while at high $\sigma$, the interface stiffness promotes surface restoration, generating intense localized shear and enstrophy, effectively acting as a source of enstrophy.

Coarse-grained analysis of the scalar variance and kinetic energy budgets revealed the mechanisms governing cross-scale transfer. The scalar variance transfer flux, $\Theta_\ell$, correlates strongly with filtered interface stretching, $\overline{\mathcal{S}}_\ell$, via a nonlinear model. During the growth stage, $\Theta_\ell$ is positive (forward cascade), driven by interface stretching, while at late times, it becomes negative (inverse transfer), driven by compression and coalescence. 
In the kinetic energy budget, three terms dominate the transfer across scales: deformation work ($\Pi_\ell$), baropycnal work ($\Lambda_\ell$), and surface tension power ($\Psi_{\sigma,\ell}$). Crucially, surface tension power is scale-dependent: it acts as a kinetic energy sink at large scales (where velocity and surface tension force are anti-aligned during breakup) and as a source at small scales (where they align during relaxation). The crossover scale between these two regimes aligns remarkably well with the classical Hinze scale definition.

By isolating individual fluid elements via additional simulations, we found that the generation of bubbles and droplets is dominated by interface stripping from the bulk, leading to a linear increase in element count over time. Morphologically, large structures are predominantly filamentary and vertically oriented, with cross-sectional diameters approximately three times the capillary scale. Consequently, the volume and surface area of these elements exhibit a linear relationship, distinct from the scaling of spherical objects. Dynamically, the element-averaged vertical velocity scales with the square root of the equivalent diameter ($\sim d^{1/2}$), consistent with a drag-buoyancy balance, while enstrophy and squared strain-rate scale linearly with size.

Finally, we verified a fundamental kinematic relationship connecting hydrodynamics to geometry: element-averaged surface tension power is directly balanced by interface stretching, $\langle \boldsymbol{f}^\sigma\cdot\bu\rangle \approx -\langle \sigma\mathcal{S}~|\nabla c|\rangle$. This relation holds across all regimes—bubbles and droplets, early and late times—and serves as the physical link between the kinetic energy budget and the scalar variance cascade. This finding holds promising implications for the development of large-eddy-simulation models for immiscible turbulence. It suggests that subgrid energy transfer to surface tension can be effectively parameterized as a function of the resolved strain rate, allowing for the accurate modeling of subgrid interface production and damping without the need to explicitly resolve complex subgrid geometries.

\begin{bmhead}[Acknowledgement]
This work is supported by the National Natural Science Foundation of China (Nos.~12202270 and 12372264) and partially by the open funds of the State Key Laboratory for Strength and Vibration of Mechanical Structures (No. SV2025-KF-02). The authors also appreciate the computational support from the Centre for High Performance Computing at Shanghai Jiao Tong University.
\end{bmhead}

\begin{bmhead}[Declaration of interests]
The authors report no conflict of interest.
\end{bmhead}

\begin{appen}

\section{Density visualizations at late time}

The late-time density visualizations at $\widehat{t}=13.6$ for cases X2, X4, and X8 are shown in figure~\ref{Appfig:late_time_viz}, providing a complementary view to the earlier-time results presented in figure~\ref{fig:field_viz}. These visualizations show that, at this later stage, the density fields in the high-$\sigma$ cases have largely reorganized into a stably stratified configuration, with substantially reduced interfacial overturning compared with the earlier snapshots.

\begin{figure}
\centering 
\begin{minipage}[b]{1.0\textwidth}  
\centering
\subfigure{\includegraphics[height=2.8in]{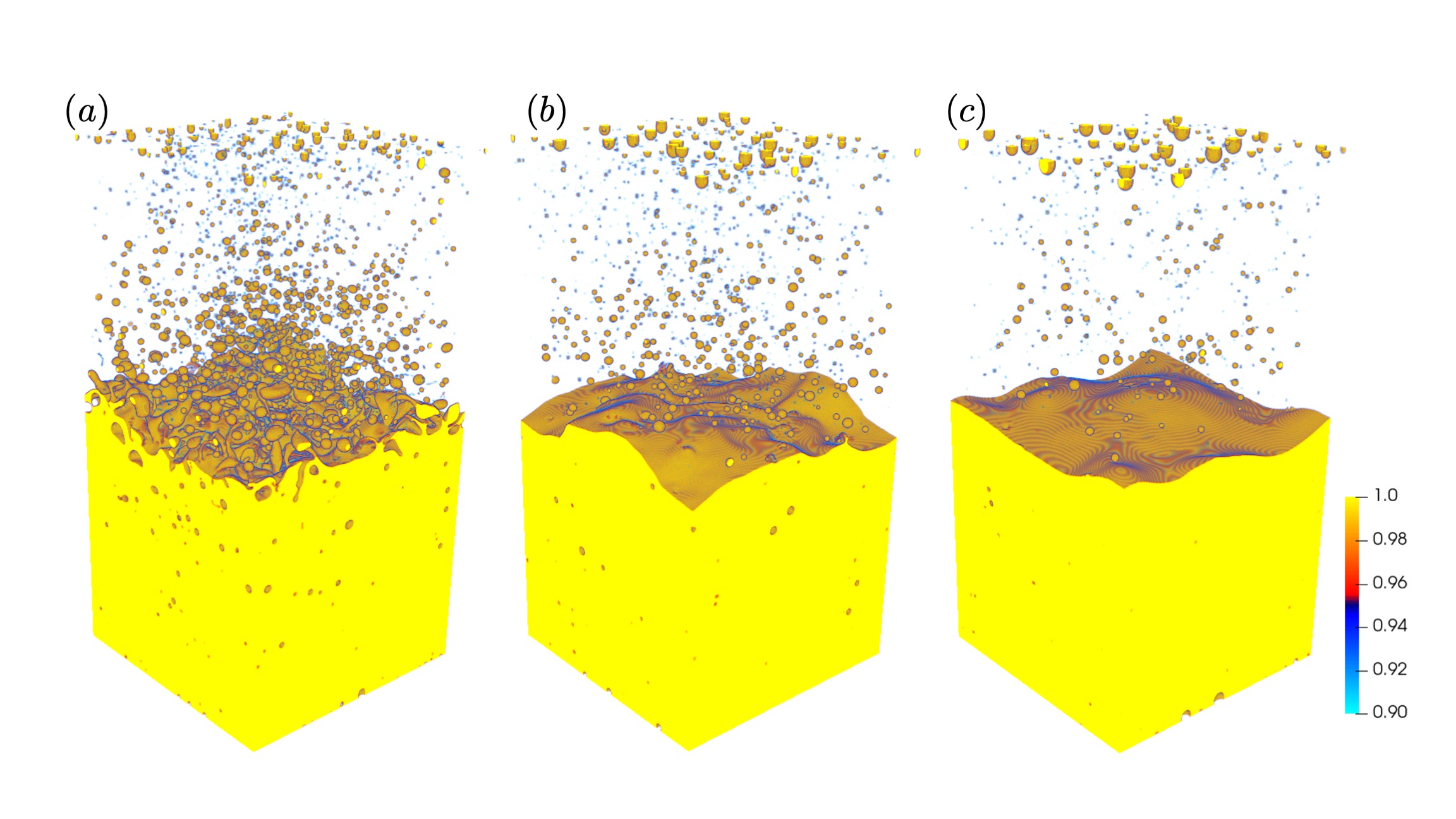} } 
\caption{Panels (a)-(c) show density visualizations for cases X2, X4, and X8, respectively, at the nondimensional time $\widehat{t}=13.6$. \label{Appfig:late_time_viz}}
\end{minipage}
\end{figure}

\section{Hinze scale in longer RT domain}

To verify the hypothesis that the minimum Hinze scale in immiscible RT turbulence is bounded by the capillary scale, we performed an additional simulation with a 4:1 aspect ratio (Case X1Ratio4), extending the parameter space beyond the 2:1 aspect ratio used in cases X1-X30. The evolution of the Hinze scale for this case is shown in Figure~\ref{Appfig:Hinze_ratio4}, plotted alongside the corresponding capillary length. We observe that the Hinze scale approaches the capillary scale from above, reaching a minimum value comparable to $l_c$ without falling below it. This behavior is consistent with the observations for the 2:1 aspect ratio cases (Figure~\ref{fig:scales}(c)), thereby further confirming the validity of our argument.

\begin{figure}
\centering 
\begin{minipage}[b]{1.0\textwidth}  
\centering
\subfigure{\includegraphics[height=2in]{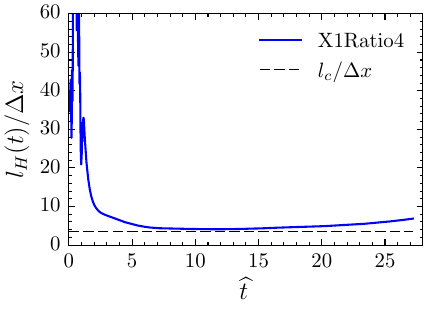} } 
\caption{Temporal evolution of the Hinze scale for case X1Ratio4 (see Table \ref{tab:parameter}). The horizontal dashed line indicates the capillary length scale, $l_c$, normalized by the grid spacing $\Delta x$. \label{Appfig:Hinze_ratio4}}
\end{minipage}
\end{figure}

\section{Density representation of the filtered scalar and kinetic-energy budgets} \label{sec:density_plot}

\begin{figure}
\centering 
\begin{minipage}[b]{1.0\textwidth}  
\centering
\subfigure
{\includegraphics[height=1.3in]{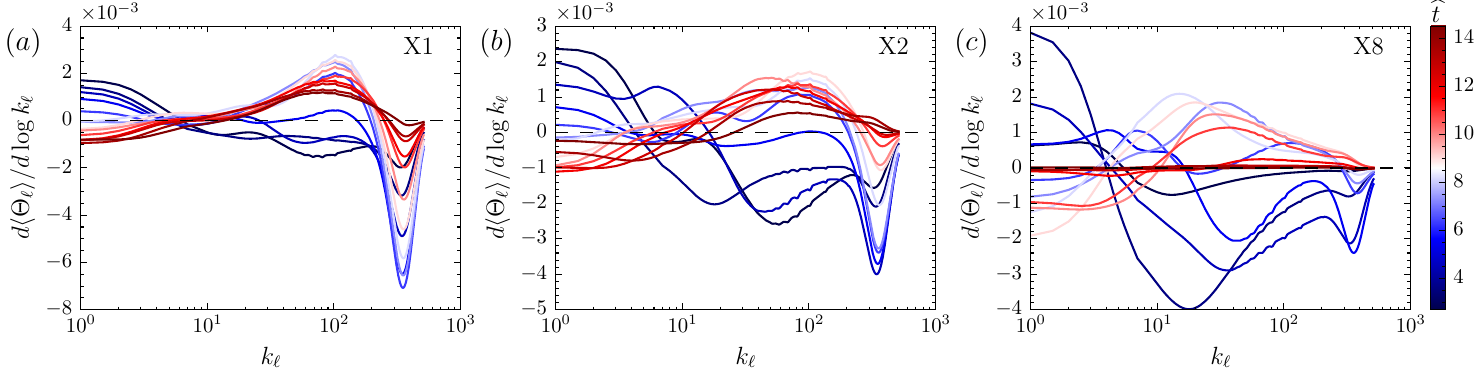} } 
\caption{Spatial average of the scalar variance transfer density $d\langle\Theta_\ell\rangle/d \log k_\ell$ versus the filtering wavenumber $k_\ell=L_x/\ell$ at different time instants indicated by the colorbar. Panels (a),(b),(c) corresponds to simulation cases X1, X2, and X8, respectively. \label{Appfig:cflux_density}}
\end{minipage}
\end{figure}

\begin{figure}
\centering 
\begin{minipage}[b]{1.0\textwidth}  
\centering
\subfigure
{\includegraphics[height=2.2in]{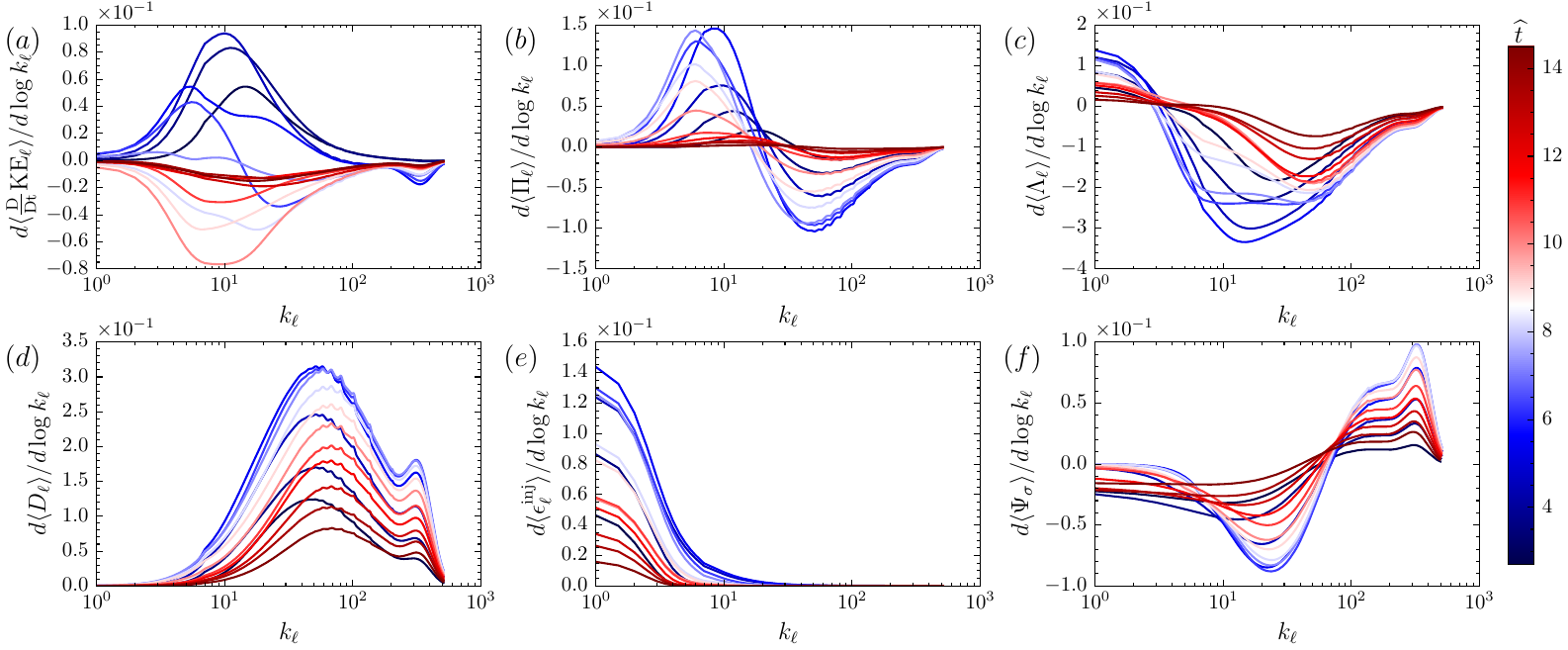} } 
\caption{Filtered kinetic-energy budgets, differentiated with respect to $\log k_\ell$, for Case X1 as a function of the filtering wavenumber $k_\ell$. Curves denote different times, color-coded by the colorbar. All terms are normalized by the maximum unfiltered injection rate over the simulation. \label{Appfig:flux_density}}
\end{minipage}
\end{figure}

To complement the budget analysis, we also compute the corresponding logarithmic-scale densities of the scalar-variance and kinetic-energy transfer budgets,
$$\frac{d(\cdot)}{d\log k_\ell}=k_\ell \frac{d(\cdot)}{d k_\ell},$$
shown in figures~\ref{Appfig:cflux_density} and~\ref{Appfig:flux_density} for the scalar-variance transfer term and the kinetic-energy budget terms, respectively. These density plots offer an alternative way to show the scale-by-scale contribution and support the same physical conclusions as the cumulative plots in figures~\ref{fig:filtered_c_var} and~\ref{fig:KE_cascade}.

\section{Compensated dissipation coefficient}

\begin{figure}
\centering 
\begin{minipage}[b]{1.0\textwidth}  
\centering
\subfigure{\includegraphics[height=2in]{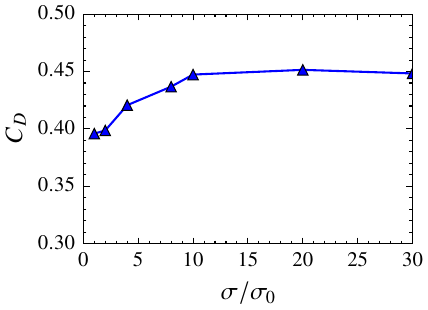} } 
\caption{Compensated coefficient $C_D$ as a function of surface tension coefficient for cases X1–X30. \label{Appfig:Cd_sigma}}
\end{minipage}
\end{figure}

The compensated dissipation coefficient is defined as
$$C_D=\frac{\max(D)\rho_m^{3/2}}
{\Delta\rho^{5/4}g^{5/4}\sigma^{1/4}},$$
where $\rho_m=(\rho_h+\rho_l)/2$ is the mean density. We have evaluated $C_D$ for cases X1–X30, as shown in figure~\ref{Appfig:Cd_sigma}. Over the range of surface tension considered, $C_D$ remains approximately constant, around 0.4-0.45. This indicates that the observed scaling, $\max(D)\propto \sigma^{1/4}$, is not merely an empirical power law, but is consistent with the full dimensional relation.

\end{appen}

\bibliographystyle{jfm}
\bibliography{refs}

\end{document}